\documentclass[%
aip,
amsmath,amssymb,
reprint,%
]{revtex4-1}

\usepackage{graphicx}
\usepackage{dcolumn}
\usepackage{bm}
\usepackage{diagbox}
\usepackage{multirow}
\usepackage{color}
\usepackage[utf8]{inputenc}
\usepackage[T1]{fontenc}
\usepackage{amsfonts}
\usepackage{amssymb}
\usepackage{amsmath}
\RequirePackage{xpatch}
\xpatchcmd\citenum{\NAT@parfalse}{\NAT@partrue}{}{}

\usepackage{lineno,hyperref}    

\newcommand{\MCC}[1]{\textcolor{black}{{#1}}}
\newcommand{\MCS}[1]{\textcolor{black}{{#1}}}
\newcommand{\CT}[1]{\textcolor{black}{{#1}}}
\newcommand{\CTS}[1]{\textcolor{black}{{#1}}}

\begin{document}
	\title{Combining stochastic density functional theory with deep potential molecular dynamics to study warm dense matter}

	\author{Tao Chen}
	\affiliation{HEDPS, CAPT, College of Engineering and School of Physics, Peking University, Beijing, 100871, P. R. China}

	\author{Qianrui Liu}
	\affiliation{HEDPS, CAPT, College of Engineering and School of Physics, Peking University, Beijing, 100871, P. R. China}

	\author{Yu Liu}
	\affiliation{HEDPS, CAPT, College of Engineering and School of Physics, Peking University, Beijing, 100871, P. R. China}

	\author{Liang Sun}
	\affiliation{HEDPS, CAPT, College of Engineering and School of Physics, Peking University, Beijing, 100871, P. R. China}

	\author{Mohan Chen}
	\thanks{Corresponding author. Email: mohanchen@pku.edu.cn}
	\affiliation{HEDPS, CAPT, College of Engineering and School of Physics, Peking University, Beijing, 100871, P. R. China}

	\date{\today}
	
	\begin{abstract}
 \MCS{
 	In traditional finite-temperature Kohn-Sham density functional theory (KSDFT), \CT{the partial occupation of a large number of high-energy KS eigenstates} restricts the use of first-principles molecular dynamics methods at extremely high temperatures. However, stochastic density functional theory (SDFT) can overcome the limitation. Recently, SDFT and its related mixed stochastic-deterministic density functional theory, based on the plane-wave basis set, have been implemented in the first-principles electronic structure software ABACUS [Phys. Rev. B 106, 125132 (2022)]. In this study, we combine SDFT with the Born-Oppenheimer molecular dynamics (BOMD) method to investigate systems with temperatures ranging from a few tens of eV to 1000 eV. Importantly, we train machine-learning-based interatomic models using the SDFT data and employ these deep potential models to simulate large-scale systems with long trajectories. Consequently, we compute and analyze the structural properties, dynamic properties, and transport coefficients of warm dense matter.
    }
	
	\end{abstract}
	
	\maketitle
	
	\section{Introduction}

	\CT{Understanding materials in extremely high-temperature conditions, such as warm dense matter (WDM) and hot dense plasma (HDP), is essential for comprehending various objects, such as planetary objects~\cite{99S-Guillot}, inertial confinement fusion~\cite{17MRE-ICF}, and high-intensity, high-energy laser pulse experiments~\cite{09NF-Moses}.}
	However, the extremely high temperatures of WDM and HDP pose significant challenges in terms of measuring their properties experimentally or predicting them theoretically~\cite{14CP-WDM}.
	After decades of combined efforts from both experimental~\cite{06RMP-Remington} and theoretical~\cite{14CP-WDM} perspectives, it has been recognized that an adequate quantum-mechanical description of electrons is essential for theoretical calculations to have the prediction power.
	In fact, various first-principles approaches based on different levels of approximations are available to address this issue.
	To name a few, the Kohn-Sham density functional theory (KSDFT)~\cite{64PR-Hohenberg, 65PR-Kohn}, the path-integral Monte Carlo (PIMC)~\cite{00L-Militzer, 01L-Militzer, 10L-Hu, 12L-Driver}, the orbital-free density functional theory (OFDFT)~\cite{11L-WangCong, 13L-White, 20MRE-Kang, 20JPCM-Qianrui}, the extended first-principles molecular dynamics (Ext-FPMD)~\cite{16PP-Ext-DFT, 21B-Liu, 22CPC-Blanchet, 22CPP-Blanchet}, and the stochastic density functional theory (SDFT)\CT{~\cite{13L-Baer, 18B-Cytter, 19WCMS-Fabian, 22ARPC-Baer, 23PRE-Sharma}} methods are most popularly employed.

\MCS{
	KSDFT with the Mermin finite-temperature density-functional approach~\cite{65PRA-Mermin}  is typically employed to compute properties of materials at low temperatures.
	However, when temperatures are elevated to the WDM regime, \CT{the partial occupation of a large number of high-energy KS eigenstates} becomes a severe hurdle as the number of electronic states that need to be considered is proportional to $T^\frac{3}{2}$, rendering the KSDFT method unfeasible\CT{~\cite{01L-Surh, 13PP-Wang, 14L-Sjostrom, 22PRR-Fiedler}}.
	In contrast, the cost of PIMC calculations decreases significantly at higher temperatures~\cite{00L-Militzer, 01L-Militzer, 10L-Hu, 12L-Driver}.
	In practice, although combining KSDFT with PIMC has been successfully applied to study the equation of states for low $Z$ elements~\cite{21PRE-Militzer}, PIMC is largely limited at lower temperatures~\cite{19PRE-Dornheim}.
	Compared with KSDFT, the OFDFT method avoids diagonalizing the wave functions and owns a higher efficiency~\cite{11L-WangCong, 13L-White, 20MRE-Kang, 20JPCM-Qianrui}.
	However, the applications of OFDFT are still limited by the lack of satisfactory accuracy in the kinetic energy density functional~\cite{18B-Luo}.
}
	
\MCS{
	The Ext-FPMD method has been proposed to evaluate systems at extremely high temperatures with the first-principles accuracy~\cite{16PP-Ext-DFT, 21B-Liu, 22CPC-Blanchet, 22CPP-Blanchet}.
	This method treats the wave functions of high-energy electrons as plane waves analytically, thereby avoiding \CT{the partial occupation of a large number of high-energy KS eigenstates}.
	However, it is still challenging to adopt the Ext-FPMD method to investigate the electrical conductivity and electronic thermal conductivity of materials due to the lack of orbital information for high-energy electrons.
	Similarly, the SDFT scheme adopts stochastic orbitals in combination with the Chebyshev trace method to overcome \CT{the partial occupation of a large number of high-energy KS eigenstates} limit.
	Moreover, the statistical errors can be effectively reduced when more stochastic orbitals are included or larger systems are utilized.
	In addition, the mixed stochastic-deterministic DFT (MDFT) combines the KS and stochastic orbitals and speeds up calculations~\cite{20L-White}.
	Note that the SDFT and MDFT methods, based on the plane-wave basis set, can be used with the $k$-point sampling method and have recently been implemented in the first-principles package ABACUS~\cite{10JPCM-ABACUS, 16CMS-ABACUS, 22B-Liu}. Since both SDFT and MDFT employ stochastic wave functions, we will refer to these two methods as SDFT throughout the remainder of this paper.
}

	\begin{figure*}[htbp]
	\begin{center}
	\includegraphics[width=17cm]{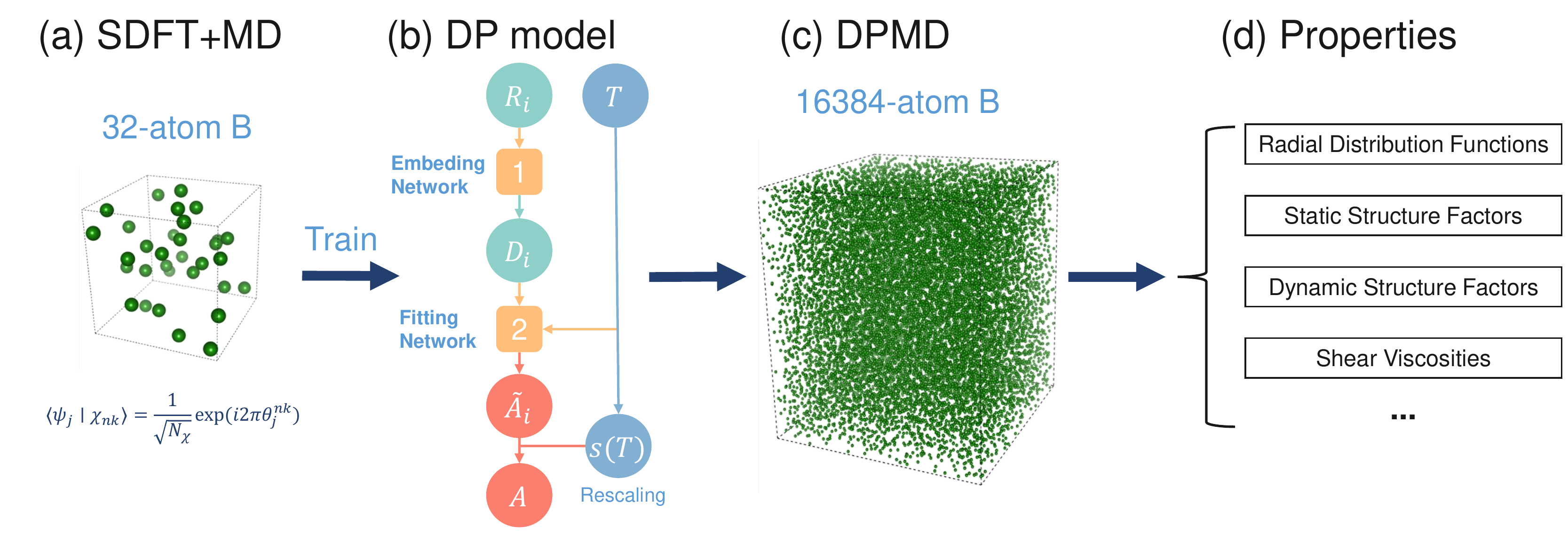}
	\end{center}
	\caption{
 \MCS{
	(Color online) Workflow of simulating WDM with the SDFT and DPMD methods.
(a) We employ stochastic orbitals in SDFT to perform MD simulations on smaller systems (32 atoms in a bulk B) and collect initial training data, which include atomic positions, energies, forces, and virial tensors. (b) We utilize the gathered training data to construct a DP model with the temperature-dependent DPMD model. The deep neural network contains both embedding and fitting networks. (c) The new model enables us to perform simulations on large systems (16384 atoms) and at extremely high temperatures (350 eV). (d) Several physical quantities such as radial distribution functions, static structure factors, dynamic structure factors, and shear viscosities can be calculated, and the data are converged with large systems and long trajectories.}
	}
	\label{fig: workflow}
	\end{figure*}
 
\MCS{
	Recently, machine-learning-assisted atomistic simulation methods have achieved remarkable success and attracted considerable attention\CT{~\cite{11PCCP-Behler, 13JPCA-Mora, 15QC-GAP, 19MP-Ko, 19NC-Pun, 19NC-Smith, 22B-dos, 20PNAS-Gartner, 22B-al, 22B-cu, 22PRM-Fiedler, 23PRR-Kumar}}.
	In particular, deep neural networks are often adopted to learn the potential energy surfaces, which are formed by the relations between atomic configurations and their corresponding energies, forces, and stresses.
	Essentially, these neural-network-based models demonstrate high efficiency while maintaining {\it ab initio} accuracy, as they effectively bypass the need to solve quantum mechanics-based equations.
	Among them, the deep potential molecular dynamics (DPMD) method~\cite{18PRL-deepmd, 18CPC-deepmd, 20PP-Yuzhi, 22MF-Wen} achieves high performances~\cite{20CPC-Denghui, 20SC-Weile} and stands out as a promising method to study WDM.
	For example, the DPMD method has been applied to study the structural and dynamic properties of warm dense aluminum~\cite{20JPCM-Qianrui, 21PRR-DaiJiayu}, the ion-ion dynamic structure factor of warm dense aluminum~\cite{21PRR-DaiJiayu}, the thermal transport by electrons and ions of warm dense aluminum~\cite{21MRE-Liu}, and the principal Hugoniot curves of warm dense beryllium~\cite{20PP-Yuzhi}, etc.
}

\MCS{
	To summarize, two main challenges exist for large-scale first-principles simulations of WDM. First, the existence of \CT{the partial occupation of a large number of high-energy KS eigenstates} results in computationally expensive simulations of WDM at high temperatures. Second, obtaining converged results for certain physical properties of WDM with a small number of atoms is difficult unless a large cell with a long MD trajectory is obtained. %
	In this work, we first validate the accuracy of SDFT by analyzing the statistical errors from the stochastic orbitals and compare with results from the traditional KSDFT method. Here, we select warm dense boron (B) and carbon (C) as benchmark systems.
	Second, we couple the Born-Oppenheimer molecular dynamics (BOMD) method with SDFT to simulate warm dense B. Specifically, we generate two DP models to describe B at a density of 2.46 $\mathrm{g/cm^3}$ with two different temperatures (86 and 350 eV); the training data are obtained from SDFT-based BOMD simulations.
	Third, by performing DPMD simulations, we significantly extend the time and spatial scales of warm dense B and obtain converged data for certain physical properties. 
    The above workflow is shown in Fig.~\ref{fig: workflow}.
  	Our work demonstrates that combining SDFT with the deep potential (DP) method offers a promising route to simulate WDM over a wide range of temperatures.}

\MCS{
	The rest of the paper is organized as follows.
	Section~\ref{Methods} describes the computational methods utilized in this work.
	In Section~\ref{Results}, we discuss the results obtained from SDFT and the DP model.
	Finally, we summarize our results in Section~\ref{Conclusions}.
 }
	
	\section{Computational Methods}
	\label{Methods}
	
	\subsection{Stochastic and Mixed Stochastic-Deterministic Density Functional Theory}
	\label{SDFT}

\MCS{
	In the KSDFT framework~\cite{64PR-Hohenberg, 65PR-Kohn}, the single-particle DFT Hamiltonian is defined as
	\begin{equation}
	\hat{H}_{\mathrm{DFT}} = -\nabla^{2} / 2+\hat{V}_{\mathrm{Hartree}}+\hat{V}_{\mathrm{xc}}+\hat{V}_{\mathrm{ext}},
	\end{equation}
	where the first term depicts the kinetic operator, $\hat{V}_{\mathrm{Hartree}}$ denotes the Hartree potential, $\hat{V}_{\mathrm{xc}}$ is the exchange-correlation (XC) potential, and $\hat{V}_{\mathrm{ext}}$ is the potential of interactions between the electrons and the nuclei as well as other external fields.
	%
	Within the finite-temperature Mermin-Kohn-Sham theory~\cite{65PRA-Mermin}, the electron density is given by
	\begin{equation}
	\begin{aligned}
	n(\mathbf{r})
	&=2 \mathrm{Tr}\left[f\left(\hat{H}_{\mathrm{DFT}},\mu,T\right)\delta(\hat{\mathbf{r}}-\mathbf{r})\right]\\
	&=2 \sum_{k=1}^{N_k}\omega_k\sum_{i=1}^{N_{\mathrm{KS}}}f\left(\varepsilon_{ik},\mu,T\right)\left|\phi_{ik}(\mathbf{r})\right|^{2},
	\end{aligned}
	\end{equation}
	where $\hat{\mathbf{r}}$ is the position operator of the electron, the prefactor 2 accounts for the electron spin. The parameter $\omega_k$ is the weight of the $k$ point with $N_k$ being the number of $k$ points, while $N_{\mathrm{KS}}$ is the number of occupied orbitals with $i$ being the index. The Fermi-Dirac distribution function takes the form of $f\left(\varepsilon_{ik},\mu,T\right)=1/[e^{(\varepsilon_{ik}-\mu)/T}+1]$, where $\mu$ is the chemical potential.
	Here, $\phi_{ik}(\mathbf{r})$ and $\varepsilon_{ik}$ respectively depict the eigenfunction and eigenvalue of the self-consistent KS Hamiltonian
	\begin{equation}
	\hat{H}_{\mathrm{DFT}} \phi_{ik}(\mathbf{r})=\varepsilon_{ik} \phi_{ik}(\mathbf{r}).
	\end{equation}
	In practice, solving $\phi_{ik}(\mathbf{r})$ and $\varepsilon_{ik}$ is costly at high temperatures since the needed number of KS wave functions is proportional to $T^\frac{3}{2}$~\cite{18B-Cytter}.
 }

\MCS{
Given any orthogonal and complete basis set $\{\psi_{j}\}$, a stochastic orbital $\chi_{nk}$ in SDFT~\cite{13L-Baer,18B-Cytter} can be defined as
	\begin{equation}
	\left\langle\psi_{j} \mid \chi_{nk}\right\rangle=\frac{1}{\sqrt{N_{\mathrm{sto}}}} \exp \left(i 2 \pi \theta_{j}^{nk}\right),
	\end{equation}
	which satisfies
	\begin{equation}
	\lim _{N_{\mathrm{sto}} \rightarrow+\infty} \sum_{n=1}^{N_{\mathrm{sto}}}\left|\chi_{nk}\right\rangle\left\langle\chi_{nk}\right|=\hat{I},
	\end{equation}
	where $\theta_{j}^{nk}$ is randomly generated by a uniform distribution between 0 and 1, and $N_{\mathrm{sto}}$ is the number of stochastic orbitals.}

 \MCS{
	In MDFT~\cite{20L-White}, both deterministic orbitals $\phi_{ik}$ and stochastic orbitals $\tilde{\chi}_{nk}$ are used, 
	\begin{equation}
\left|\tilde{\chi}_{nk}\right\rangle=\left|\chi_{nk}\right\rangle-\sum_{i=1}^{N_{\mathrm{KS}}}\left\langle\phi_{ik} \mid \chi_{nk}\right\rangle\left|\phi_{ik}\right\rangle.
	\end{equation}
	Here, the stochastic orbitals are defined to be orthogonal to the deterministic orbitals. The number of deterministic orbitals is set to $N_{\mathrm{KS}}$, which is typically chosen to be a subset of occupied states.
 In addition, both sets of orbitals satisfy the relation
	\begin{equation}
	\lim _{N_{\mathrm{sto}} \rightarrow+\infty} \sum_{n=1}^{N_{\mathrm{sto}}}\left|\tilde{\chi}_{nk}\right\rangle\left\langle\tilde{\chi}_{nk}\right|+\sum_{i=1}^{N_{\mathrm{KS}}}\left|\phi_{ik}\right\rangle\left\langle\phi_{ik}\right|=\hat{I},
	\end{equation}
	Then, the electron density is given by
	\begin{equation}
	\begin{aligned}
n(\mathbf{r})=2\sum_{k=1}^{N_k}\omega_k\Bigg[\sum_{n=1}^{N_{\mathrm{sto}}}\left|\left\langle\mathbf{r}\left|\sqrt{f\left(\hat{H}_{\mathrm{DFT}},\mu,T\right)}\right| \chi_{nk}\right\rangle\right|^{2}\\
+\sum_{i=1}^{N_{\mathrm{KS}}}f\left(\varepsilon_{ik},\mu,T\right)\left|\phi_{ik}(\mathbf{r})\right|^{2}\Bigg],
	\end{aligned}
	\end{equation}
	where $\sqrt{f\left(\hat{H}_{\mathrm{DFT}},\mu,T\right)}$ is calculated by the Chebychev expansion~\cite{97L-Baer}.
	If $N_{\mathrm{sto}}=0$ or $N_{\mathrm{KS}}=0$, the MDFT method changes to the standard KSDFT or SDFT methods, respectively.
 }
	
\MCS{
 	Notably, using stochastic orbitals in practical calculations results in statistical errors since these orbitals only form a complete basis when the number of stochastic orbitals approaches infinity. 
  	As reported in previous studies~\cite{19-WCMS-Fabian, 22ARPC-Baer, 22B-Liu}, the error caused by the stochastic orbitals is proportional to $1/\sqrt{N_{\mathrm{sto}}}$.
  	However, when periodic boundary conditions (PBCs) with the $k$-point sampling are considered and each $k$-point has $N_{\mathrm{sto}}$ stochastic orbitals, the resulting $\sigma_s$ is proportional to $1/\sqrt{N_k N_{\mathrm{sto}}}$, suggesting that more $k$-points can reduce the stochastic errors. In order to evaluate the accuracy of SDFT, we perform both SDFT and KSDFT calculations for B and C bulk systems in extreme conditions.
   }

 \MCS{
	In the first-principles MD simulations of WDM, the KSDFT couples with the dynamics of ions, usually through the BOMD method.
	Since the motions of ions are treated classically in the BOMD method, we need to evaluate the force of an atom $i$ in the form of
	\begin{equation}
	\mathbf{F}_{i}=-\frac{\partial E}{\partial \mathbf{R}_{i}}.
	\end{equation} 
    Here $E$ is the \CT{sum of electrons' energy and ion-ion repulsion energy}, and $\mathbf{R}_{i}$ is the position of atom $i$.
	By utilizing the plane-wave basis set and norm-conserving pseudopotentials within the traditional KSDFT and SDFT methods, the force of an atom can be decomposed into three parts
	\begin{equation}
\mathbf{F}_{i}=\mathbf{F}_{i}^{L}+\mathbf{F}_{i}^{NL}+\mathbf{F}_{i}^{II},
	\end{equation}
	where $\mathbf{F}_{i}^{L}$ is the local pseudopotential force term, $\mathbf{F}_{i}^{NL}$ depicts the nonlocal pseudopotential force term, and $\mathbf{F}_{i}^{II}$ is the Ewald force term origins from the ion-ion interactions.
 	Furthermore, stress is defined as
  	\begin{equation}
  	\begin{aligned}
  	\sigma_{\alpha \beta}&=-\frac{1}{V} \frac{\partial E}{\partial \epsilon_{\alpha \beta}}\\
  	&=
   \sigma_{\alpha \beta}^{T}
   +\sigma_{\alpha \beta}^L
   +\sigma_{\alpha \beta}^{NL}
   +\sigma_{\alpha \beta}^{\text {Hartree}}
   +\sigma_{\alpha \beta}^{\mathrm{xc}}
   +\sigma_{\alpha \beta}^{II},
 	\end{aligned}
  	\end{equation}
   	where $\epsilon_{\alpha \beta}$ is the strain with the spatial coordinates $\alpha$ and $\beta$. 
    The kinetic energy term of electrons is $\sigma_{\alpha \beta}^T$. 
    $\sigma_{\alpha \beta}^L$ and $\sigma_{\alpha \beta}^{NL}$ are the local and nonlocal pseudopotential terms, respectively.
$\sigma_{\alpha \beta}^{\text {Hartree }}$ is the Hartree and $\sigma_{\alpha \beta}^{\mathrm{xc}}$ is the exchange-correlation term. The Ewald term is $\sigma_{\alpha \beta}^{II}$.
	One can refer to Ref.~\onlinecite{22B-Liu} by Liu {\it et al.} for more details on implementing  \CT{the total energy, the total free energy,} forces, and stresses within the framework of SDFT in ABACUS.}
    \CT{Note that the SDFT method employed in this work still uses plane wave basis, so it is still computationally demanding for a system consisting of a few hundred atoms.}


	\subsection{Deep Potential Molecular Dynamics}
	\label{DPMD}
 
	\MCS{
	In the DPMD method~\cite{18L-deepmd, 18CPC-deepmd}, the total energy $E$ of a system is expressed as a sum of atomic contributions, i.e., $E=\sum_i E_i$, where the energy $E_i$ from atom $i$ depends on an environment matrix $\mathcal{R}_i$, which includes the information of neighboring atoms of atom $i$ within a cutoff radius.
	The DP model maps $\mathcal{R}_i$ via an embedding neural network to a symmetry-preserving descriptor, and then the descriptor is mapped to a fitting neural network to yield $E_i$.
	A loss function is utilized to optimize the parameters in the embedding and fitting networks to generate DP models. The loss function is defined as
	\begin{equation}
L(p_\epsilon,p_f,p_\xi)=p_\epsilon\Delta\epsilon^2+\frac{p_f}{3N}\sum_{i}|\Delta\mathbf{F}_i|^2+\frac{p_\xi}{9}||\Delta\xi||^2,
	\end{equation}
	where $N$ is the number of atoms,
	$\epsilon = E/N$ is the energy per atom,
	$\mathbf{F}_i$ is the force acting on atom $i$,
	$\xi$ is the virial tensor per atom, and
	$\Delta$ denotes the difference between the training data and the results predicted by the DP model.
	In addition, $p_\epsilon$, $p_f$, and $p_\xi$ are tunable prefactors.
	The stochastic gradient descent scheme Adam~\cite{14arX-Adam} is adopted to train the DP model.
}

\MCS{
	We separately train the data from SDFT-based BOMD trajectories of B at 86 and 350 eV, where the Perdew-Burke-Ernzerhof (PBE)~\cite{96L-PBE} functional is used. As a result, we obtain two DP models of warm dense B at the two temperatures.
	To better characterize the warm dense B at 350 eV, we adopt the temperature-dependent deep potential (TDDP) method~\cite{20PP-Yuzhi} to train the DP model.
	Note that the TDDP method, as shown in Fig.~\ref{fig: workflow}(b), introduces the electron temperature of the system into the fitting net, which is more suitable for high-temperature systems.
 }

 \MCS{
	Both embedding and fitting neural networks contain three layers with the specific number of neurons being (25, 50, 100) and (120, 120, 120), respectively.
	The cutoff radius for each atom is chosen to be 6.0 {\AA}.
	The inverse distance $1/r$ decays smoothly from 0.5 to 6.0 {\AA} in order to remove the discontinuity introduced by the cutoff.
	Both DP models undergo training for 500,000 steps. Throughout the training process, the values of $p_\epsilon$, $p_f$, and $p_\xi$ are gradually adjusted from 0.02 to 1, 1000 to 1, and 0.02 to 1, respectively.
 	We also employ the DP compress technique to both DP models to accelerate the DPMD simulations, as described in the literature~\cite{22JCTC-Lu}.
}

 \section{Results and Discussion}
 \label{Results}
	
 \subsection{Statistical Errors of SDFT}

  	\begin{figure}[htbp]
	\begin{center}
	\includegraphics[width=8.5cm]{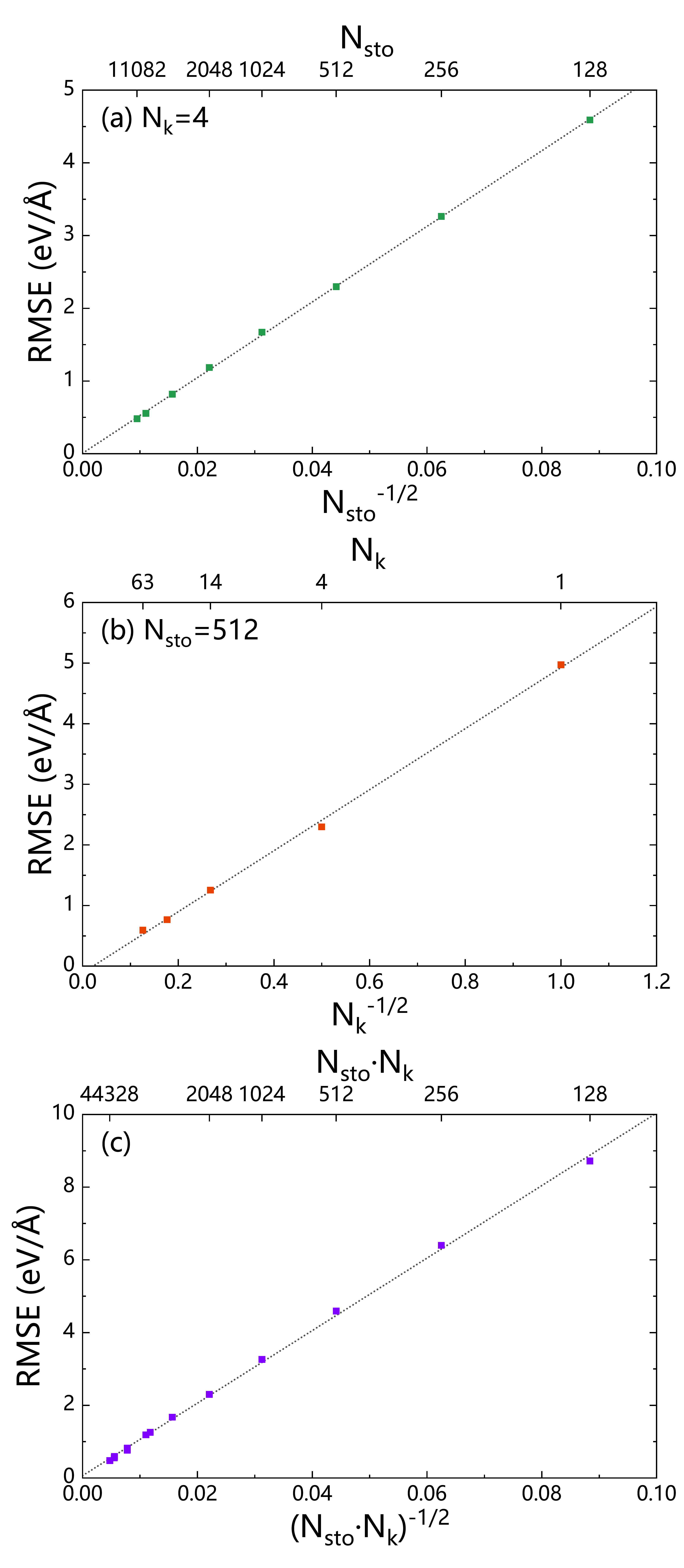}
	\end{center}
	\caption{\MCS{
	(Color online) Root-mean-square error (RMSE) of atomic forces arises from the SDFT calculations for B. The temperature is set to 350 eV and the density is set to 2.46 $\mathrm{g/cm^3}$. The RMSE is evaluated with respect to (a) the number of stochastic orbitals $N_{\mathrm{sto}}$, (b) the number of $k$ points in the Brillouin zone $N_{k}$, and (c) the product of $N_{\mathrm{sto}}$ and $N_{k}$.
	For each data point, the RMSE is obtained via 9 independent SDFT calculations with different sets of stochastic orbitals.}
	}
	\label{fig: force-err}
	\end{figure}
 	


 \MCS{
 To analyze the statistical errors that arise from the SDFT method itself, we choose a 32-atom B system with a density of 2.46 $\mathrm{g/cm^3}$ at the temperature of 350 eV. In addition, we employ the PBE~\cite{96L-PBE} exchange-correlation functional.
 We note that at the temperature of 86 eV and higher temperatures, the pseudopotential of B is generated by the ONCVPSP~\cite{13B-Hamann} method with all of its 5 electrons. The cutoff radius for the pseudopotential is set to 0.7 Bohr to avoid overlaps of electron orbitals at high temperatures.
 In addition, we select an energy cutoff of 180, 240, and 300 Ry for temperatures of 86, 350, and 1000 eV, respectively.
}

\MCS{
 Figure~\ref{fig: force-err} shows the RMSE of SDFT verse the number of stochastic orbitals ($N_{\mathrm{sto}}$) and the number of $k$-point ($N_{k}$).
 For each data point in this figure, we label the average atomic force for each atom along a certain direction ($\gamma\in{x,y,z}$) as $\mathrm{F_{ave}}$, which is computed by averaging 9 independent SDFT calculations with different sets of stochastic orbitals.
 In each SDFT calculation, we label the force acting on each atom along the $\gamma$ direction as $\mathrm{F_{sto}}$. 
 In this regard, the root-mean-square error (RMSE) can be evaluated via
 \begin{equation}
\mathrm{RMSE} = \sqrt{\frac{1}{3N_c\cdot N} \sum_{c=1}^{N_c}\sum_{i=1}^{N}\sum_{\gamma} (\mathrm{F}_{\mathrm{sto}}-\mathrm{F}_\mathrm{ave})^2},
\label{eq: rmse}
 \end{equation}
 where $N_c=9$ is the number of independent SDFT runs, and $N=32$ is the number of atoms with $i$ being the index of atoms.
}

\MCS{
 The number of stochastic orbitals is chosen from 128 to 11082 in Fig.~\ref{fig: force-err}(a), and the shifted $k$-point sampling is set to 2$\times$2$\times$2. Note that after symmetry analysis, the number of $k$ points reduces to 4 after symmetry analysis.
 We fix the number of stochastic orbitals to be 512 in Fig.~\ref{fig: force-err}(b) and choose the shifted $k$-point samplings of $1\times1\times1$ (1), $2\times2\times2$ (4), $3\times3\times3$ 32), and $5\times5\times5$ (65); here the number in the parentheses denotes the number of $k$-points after symmetry analysis.
 As expected, Figs.~\ref{fig: force-err}(a), (b), and (c) respectively show that the RMSE of forces acting on atoms exhibits linear behavior with respect to $N_\mathrm{{sto}}^{-1/2}$, $N_k^{-1/2}$, and $(N_\mathrm{{sto}}\times N_k)^{-1/2}$.
 The numerical results are consistent with the discussion of statistical error in Section~\ref{SDFT}.
 Importantly, the results indicate that as more stochastic orbitals and a larger number of $k$-point sampling are employed, the stochastic errors can be systematically mitigated.
}

 \subsection{Compare SDFT and KSDFT Results}
 
 	\begin{figure*}[htbp]
	\begin{center}
	\includegraphics[width=17cm]{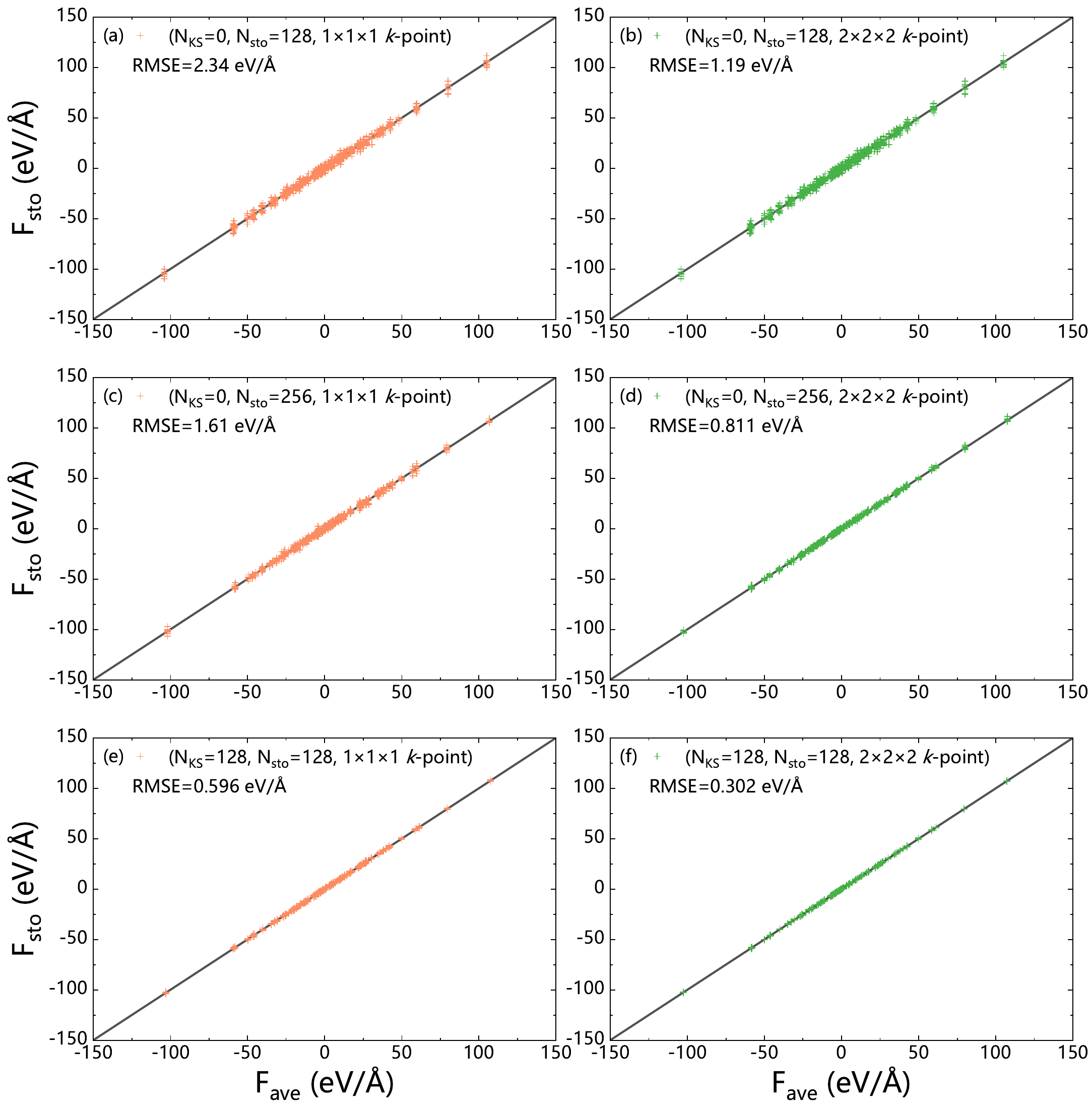}
	\end{center}
	\caption{
   \MCS{
	(Color online) Forces acting on each B atom as obtained from 9 independent SDFT calculations with different stochastic orbitals. The B system has a density of 2.46 $\mathrm{g/cm^3}$, and the temperature is 17.23 eV. For each calculation, the force acting on each atom along the $\gamma$ direction is denoted as $\mathrm{F_{sto}}$, and their average is $\mathrm{F_{ave}}$.
    $N_{\mathrm{{KS}}}$ refers to the number of KS orbitals, and $N_{\mathrm{{sto}}}$ is the number of stochastic orbitals. 
    Two sets of Monkhorst-Pack $k$-points are utilized, i.e., 
    a $1\times1\times1\ k$-point mesh (the $\Gamma\ k$-point) and a shifted $2\times2\times2\ k$-point mesh.
    The root-mean-square error (RMSE) of forces between $\mathrm{F_{sto}}$ and $\mathrm{F_{ave}}$ is computed via Eq.~\ref{eq: rmse}. The RMSE is obtained via the mentioned 9 independent SDFT calculations.}
	}
	\label{fig: 17eV}
	\end{figure*}

	\begin{figure}[htbp]
	\begin{center}
	\includegraphics[width=8.5cm]{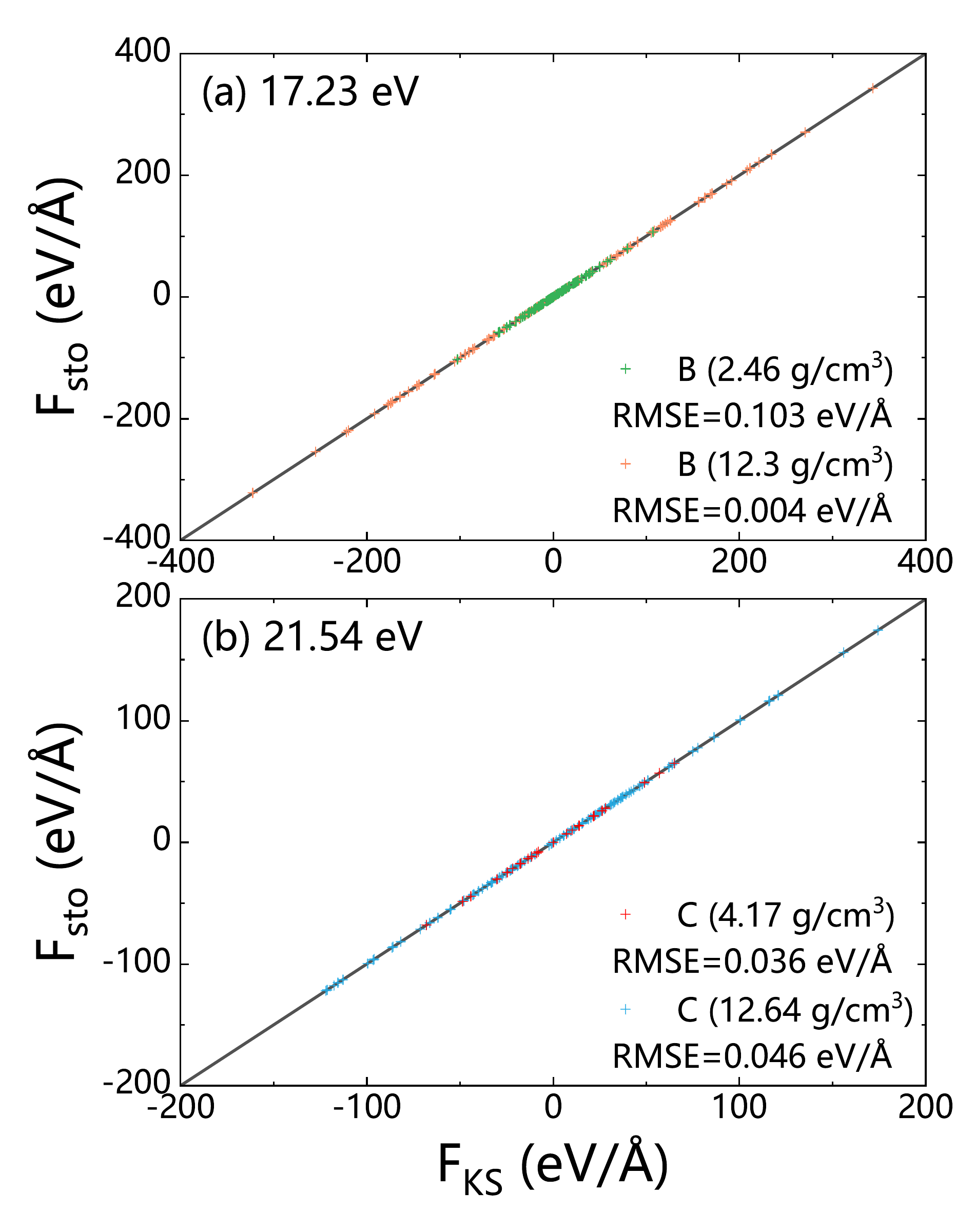}
	\end{center}
	\caption{\MCS{
	(Color online) (a) Comparison of forces acting on each B atom in a 32-atom cell with densities being 2.46 and 12.3 $\mathrm{g/cm^3}$ at a temperature of 17.23 eV. (b) Comparison of forces acting on each C atom with densities being 4.17 and 12.46 $\mathrm{g/cm^3}$ at a temperature of 21.54 eV. In the SDFT (KSDFT) calculation, we label the force acting on each atom along the $\gamma$ direction ($\gamma\in{x,y,z}$) as $\mathrm{F_{sto}}$ ($\mathrm{F_{KS}}$). The root-mean-square error (RMSE) of forces between $\mathrm{F_{sto}}$ and $\mathrm{F_{KS}}$ is listed. }
	}
	\label{fig: force-BC}
	\end{figure}
	
	\begin{figure*}[htbp]
	\begin{center}
	\includegraphics[width=17cm]{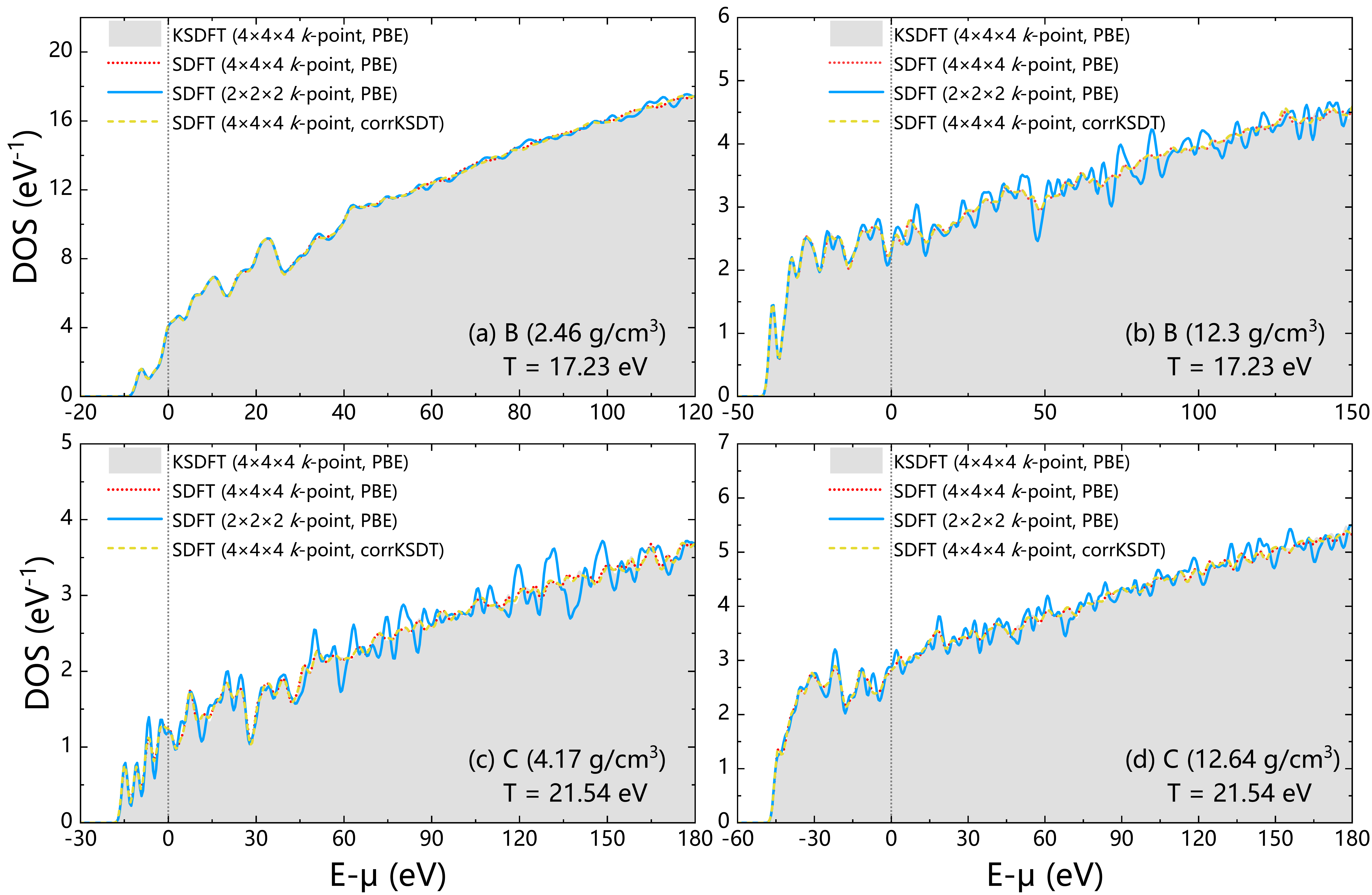}
	\end{center}
	\caption{
 \MCS{
 (Color online) Density of states (DOS) for the B and C systems as calculated by the SDFT and KSDFT methods.
 The densities are selected as (a) 2.46 $\mathrm{g/cm^3}$ and (b) 12.3 $\mathrm{g/cm^3}$ for B, and (c) 4.17 $\mathrm{g/cm^3}$ and (d) 12.64 $\mathrm{g/cm^3}$ for C systems.
 The Fermi energy is set to 0.
 We use two sets of $k$-point sampling in KSDFT and SDFT calculations. In addition, we adopt two exchange-correlation functionals, including PBE~\cite{96L-PBE} and corrKSDT~\cite{18L-corrKSDT}.
 }
	}
	\label{fig: dos}
	\end{figure*}
 
	\begin{table}[htbp]
	\centering
	\caption{ \MCS{
  Comparison of the total energy per atom ($E$ in eV), the pressure ($P$ in GPa), and the degree of ionization ($\alpha$) for B and C systems as obtained from the SDFT and the traditional KSDFT methods.
  Four systems are chosen, i.e., two B systems at a temperature of 17.23 eV with densities being 2.46 and 12.3 $\mathrm{g/cm^3}$, and two C systems with densities being 4.17 and 12.46 $\mathrm{g/cm^3}$ at a temperature of 21.54 eV. $\Delta$ denotes the percentage difference between the results obtained by SDT and KSDFT.}
	}
	\setlength{\tabcolsep}{4pt}
	\renewcommand\arraystretch{1.2}
    \begin{tabular}{c|c|ccc}
    \hline
        ~ & ~ & $E$ (eV) & $P$ (GPa) & $\alpha$ \\ \hline
        B & KSDFT & -153.683171 & 849.112 & 0.476617 \\
        2.46 $\mathrm{g/cm^3}$ & SDFT & -153.707884 & 852.524 & 0.476659 \\
        17.23 eV & $\Delta$ & 0.0161\% & 0.4019\% & 0.0088\% \\ \hline
        B  & KSDFT & -66.923177 & 8730.010 & 0.390341 \\ 
        12.3 $\mathrm{g/cm^3}$ & SDFT & -66.924235 & 8730.829 & 0.390344 \\ 
        17.23 eV & $\Delta$ & 0.0016\% & 0.0094\% & 0.0008\% \\ \hline
        C & KSDFT & -263.586543 & 2018.380 & 0.505693 \\ 
        4.17 $\mathrm{g/cm^3}$ & SDFT & -263.595810 & 2019.780 & 0.505707 \\ 
        21.54 eV & $\Delta$ & 0.0035\% & 0.0693\% & 0.0028\% \\ \hline
        C  & KSDFT & -190.692843 & 9168.050 & 0.440747 \\
        12.46 $\mathrm{g/cm^3}$ & SDFT & -190.698535 & 9171.552 & 0.440757 \\
        21.54 eV & $\Delta$ & 0.0030\% & 0.0382\% & 0.0023\% \\ \hline
    \end{tabular}
	\label{table: B_table}
	\end{table}

 \MCS{In order to select suitable numbers of $k$-points, KS orbitals, and stochastic orbitals for simulating warm dense matter under specific conditions, we use a 32-atom B system as an example, with a density of 2.46 $\mathrm{g/cm^3}$ and a temperature of 17.23 eV.
 Figure~\ref{fig: 17eV} compares the forces on each B atom with different values for the mentioned parameters.
 First, we find that increasing the $k$-point sampling size from the $\Gamma$ point to a shifted $2\times2\times2$ $k$-point sampling substantially reduces the RMSE, which is consistent across various values of $N_{\mathrm{KS}}$ and $N_{\mathrm{sto}}$.
 The result is in line with the linear relationship shown in Fig.~\ref{fig: force-err}.
 Second, we note that the RMSE with $N_{\mathrm{KS}}=0$ and $N_{\mathrm{sto}}=256$ in Fig.~\ref{fig: 17eV}(c) is 1.61 eV/~\AA~, while the RMSE with $N_{\mathrm{KS}}=128$ and $N_{\mathrm{sto}}=128$ shown in Fig.~\ref{fig: 17eV}(e) is 0.596 eV/~\AA~. The former is considerably larger than the latter, suggesting that increasing the number of KS orbitals is more effective than using stochastic orbitals at a relatively low temperature (17.23 eV).
 Consequently, by choosing an adequate number of KS orbitals ($N_{\mathrm{KS}}=128$) and stochastic orbitals ($N_{\mathrm{sto}}=128$) along with the shifted $2\times2\times2$ $k$-point sampling, we can achieve an RMSE as small as 0.302 eV/~\AA~. 
 Furthermore, we examine the effects of these parameters on B systems at 86 and 350 eV, with the results shown in Figs. S1 and S2 of Supporting Information (SI), respectively. In conclusion, we find it reasonable to select the same parameters as in the B system at 17.23 eV.
}

\MCS{
 Next, Table~\ref{table: B_table} presents a comparison of some key physical properties obtained from the SDFT and KSDFT methods. Specifically, we evaluate the total energy per atom ($E$), the pressure ($P$), and the degree of ionization ($\alpha$). $\Delta$ is the percentage difference between the results obtained from SDFT and the traditional KSDFT. We consider four systems, including two B systems at a temperature of 17.23 eV and densities of 2.46 and 12.3 $\mathrm{g/cm^3}$, as well as two C systems at a temperature of 21.54 eV and densities of 4.17 and 12.46 $\mathrm{g/cm^3}$. In both SDFT and KSDFT calculations, we choose the PBE~\cite{96L-PBE} functional. Furthermore, a shifted $2\times2\times2$ $k$-point sampling grid is adopted.
}

\MCS{
 We study B systems with a cell containing 32 atoms. Additionally, we employ a norm-conserving pseudopotential for B with 3 valence electrons~\cite{15CPC-Martin}, and the energy cutoff is 150 Ry. 
 We respectively set the number of deterministic orbitals in KSDFT to be $N_{\mathrm{KS}}=992$ and 400 for the B systems with density being 2.46 and 12.3 $\mathrm{g/cm^3}$. This setting ensures the occupation of electrons to be smaller than $10^{-4}$ at the highest-energy orbital.
 On the other hand, we choose the number of deterministic orbitals to be $N_{\mathrm{KS}}=240$ and the number of stochastic orbitals to be $N_{\mathrm{sto}}=120$ in SDFT for the B systems regardless of their densities.
 Regarding the C systems, a norm-conserving pseudopotential with 4 valence electrons is employed~\cite{15CPC-Martin}, and the energy cutoff is set to 160 Ry.
 We utilize 8 (32) atoms in a cell and $N_{\mathrm{KS}}=350\ (520)$ for a density of 4.17 (12.46) $\mathrm{g/cm^3}$ in KSDFT calculations.
 In SDFT calculations, we adopt $N_{\mathrm{KS}}=120\ (240)$ and $N_{\mathrm{sto}}=120\ (120)$ for 4.17 (12.46) $\mathrm{g/cm^3}$.}


 \MCS{
 As shown in Table~\ref{table: B_table},
 we have the following findings.
 First, it can be seen that the percentage difference in total energy ($\Delta$ of $E$) between SDFT and KSDFT is relatively small, being less than 0.02\% for the B systems and 0.004\% for the C systems. This indicates that SDFT provides a high-accuracy estimation of total energy when compared to the conventional KSDFT method.
 Second, the percentage difference in pressure ($\Delta$ of $P$) between the two methods is smaller than 0.41\% for B and 0.07\% for C. This further supports the high accuracy of the SDFT method.
 Third, the ionization process of electrons plays a crucial role in determining the WDM equation of state~\cite{16B-GaoChang, 18PRE-ZhangShuai, 22CPP-Blanchet}. This process can be represented by the degree of ionization, denoted as $\alpha$.
 In practice, the Fermi energy of the system at 0 K is defined as $\mu$. Consequently, the degree of ionization $\alpha$ at a finite temperature $T$ can be defined as follows
  \begin{equation}
  \alpha=1-\frac{N_{T,occ}}{N_{0,occ}},
  \end{equation}
  where $N_{T,occ}$ is the total number of occupied electrons below $\mu$ when the electrons follow the Fermi-Dirac distribution at tempearture $T$. In addition,
  $N_{0,occ}$ is a special case of $N_{T,occ}$ when $T=0$.
 The percentage difference in the degree of ionization ($\Delta$ of $\alpha$) is found to be smaller than 0.009\% for both B and C systems.
 In summary, all three properties, namely $E$, $P$, and $\alpha$ calculated by SDFT show excellent accuracy when compared to those from the traditional KSDFT. This demonstrates that SDFT is a reliable method for simulating high-temperature materials with first-principles accuracy.
 }

 \MCS{
 Fig.~\ref{fig: force-BC} further compares the forces acting on each atom of B (2.46 and 12.3 $\mathrm{g/cm^3}$) and C (4.17 and 12.46 $\mathrm{g/cm^3}$) obtained from both SDFT and KSDFT calculations.
 We find that the forces predicted by SDFT are in excellent agreement with those from KSDFT. For instance, the RMSE of forces is smaller than 0.05 eV/\AA~for both C systems. The largest RMSE occurs in the B system at 2.46 $\mathrm{g/cm^3}$, with a value of 0.103 eV/\AA, which is relatively small compared to the magnitude of atomic forces (a few hundreds of eV/\AA). Notably, we find the smallest RMSE is 0.004 eV/\AA~in the B system at 12.3 $\mathrm{g/cm^3}$. This is due to the fact that more electronic states of B are occupied by electrons at this condition, as demonstrated by the smaller degree of ionization of B (0.39) shown in Table.~\ref{table: B_table}.
 }

 \MCS{
 Fig.~\ref{fig: dos} illustrates the DOS of B (2.46 and 12.3 $\mathrm{g/cm^3}$) and C (4.17 and 12.46 $\mathrm{g/cm^3}$).
 Besides the PBE~\cite{96L-PBE} exchange-correlation functional, we also test the finite-temperature local density approximation functional, i.e., the corrKSDT functional as proposed by Karasiev {\it et al.}~\cite{18L-corrKSDT}.
 A Monkhorst-Pack $4\times4\times4$ shifted $k$-point mesh is adopted in KSDFT calculations to yield the DOS of B and C. However,
 unlike the traditional KSDFT method, DOS in SDFT cannot be directly obtained from the eigenvalues of $\hat{H}$. Instead, we evaluate the DOS from the SDFT method via the following formula
 \begin{equation}
  g(E)=2\operatorname{Tr}\left[\frac{1}{\sqrt{2 \pi} \sigma} \exp \left(-\frac{(E-\hat{H})^{2}}{2 \sigma^{2}}\right)\right].
 \end{equation}
 Here, $\sigma$ depicts the width of smearing.
 The DOS of SDFT with a shifted $2\times2\times2$ $k$-point mesh converges for B with a density of 2.46 $\mathrm{g/cm^3}$ when compared with KSDFT, although there are some deviations observed in the other three cases.
 Notably, the DOS predicted by SDFT using a $4\times4\times4$ shifted $k$-point mesh shows excellent agreement with the KSDFT results for both B and C systems.
 By employing a shifted $4\times4\times4$ $k$-point mesh, it is also observed that the DOS of corrKSDT~\cite{18L-corrKSDT} exhibits no significant differences when compared with the PBE~\cite{96L-PBE} results, suggesting that the temperature effects in the XC functional have minimal impacts on our calculations.
 Overall, these findings indicate that the SDFT implemented in ABACUS is adequately accurate for simulating warm dense B and C systems.
 }

\subsection{High-Temperature Calculations by SDFT}


\begin{table}[htbp]
    \centering
    \caption{ 
    \MCS{
    Root-Mean-Square Error (RMSE) of forces acting on B atoms as obtained from 9 independent SDFT calculations with two sets of stochastic orbitals (128 and 256). 
    The B system has a density of 2.46 $\mathrm{g/cm^3}$, and the temperatures are set to 17.23, 86, 350, and 1000 eV.
    $N_{\mathrm{{KS}}}$ refers to the number of KS orbitals, and $N_{\mathrm{{sto}}}$ is the number of stochastic orbitals. 
    Two sets of Monkhorst-Pack $k$-points are utilized, i.e., a $1\times1\times1\ k$-point mesh (include the $\Gamma\ k$-point) and a shifted $2\times2\times2\ k$-point mesh.
%
    The RMSE of forces is computed via Eq.~\ref{eq: rmse}. 
    }
	}
    \setlength{\tabcolsep}{8pt}
    \renewcommand\arraystretch{1.4}
    \begin{tabular}{ccccc}
    \hline
        T (eV) & $N_{\mathrm{KS}}$ & $N_{\mathrm{sto}}$ & $k$ points & RMSE (eV/Å) \\ \hline
        17.23 & 128 & 128 & $1\times1\times1$ & 0.596 \\ 
         & 128 & 128 & $2\times2\times2$ & 0.302 \\ \hline
        86 & 128 & 128 & $1\times1\times1$ & 3.26  \\ 
         & 128 & 128 & $2\times2\times2$ & 1.54  \\ \hline
        350 & 0 & 256 & $1\times1\times1$ & 6.40  \\ 
         & 0 & 256 & $2\times2\times2$ & 3.26  \\ \hline
        1000 & 0 & 256 & $1\times1\times1$ & 2.68  \\ 
         & 0 & 256 & $2\times2\times2$ & 1.30 \\ \hline
    \end{tabular}
    \label{table: force}
\end{table}

\MCS{
 Table~\ref{table: force} collects the three force components of each atom in the 32-atom B cell from 9 independent runs of SDFT with different stochastic orbitals. Four different temperatures are chosen, i.e., 17.23, 86, 350, and 1000 eV. Furthermore, we select $1\times1\times1$ and $2\times2\times2$ shifted $k$-point samplings for each temperature and evaluate the corresponding RMSE. 
 At each condition, a set of average forces $\mathrm{F}_{\mathrm{ave}} $ are calculated according to Eq.~\ref{eq: rmse}.
 One can refer to Fig. S3 of SI for more details.
 For temperatures of 17.23 and 86 eV, we set $N_{\mathrm{KS}}=128$ and $N_{\mathrm{sto}}=128$; at higher temperatures of 350 and 1000 eV, we find it more effective to use stochastic orbitals than the KS orbitals.
 As a result, we do not choose the Kohn-Sham orbitals ($N_{\mathrm{KS}}=0$) but set all orbitals to be stochastic orbitals ($N_{\mathrm{sto}}=256$). 
 According to our tests, the RMSE of the atomic force smaller than 3.3 eV/\AA\ is enough for FPMD simulations of WDM B.
 Therefore, we employ the $\Gamma\ k$-point with 128 KS orbitals and 128 stochastic orbitals for 86 eV and the $2\times2\times2$\ shifted $k$-point with 256 stochastic orbitals for 350 eV to perform FPMD.
 } 
 
\subsection{Radial Distribution Functions}

\begin{figure}[htbp]
\begin{center}
\includegraphics[width=8.5cm]{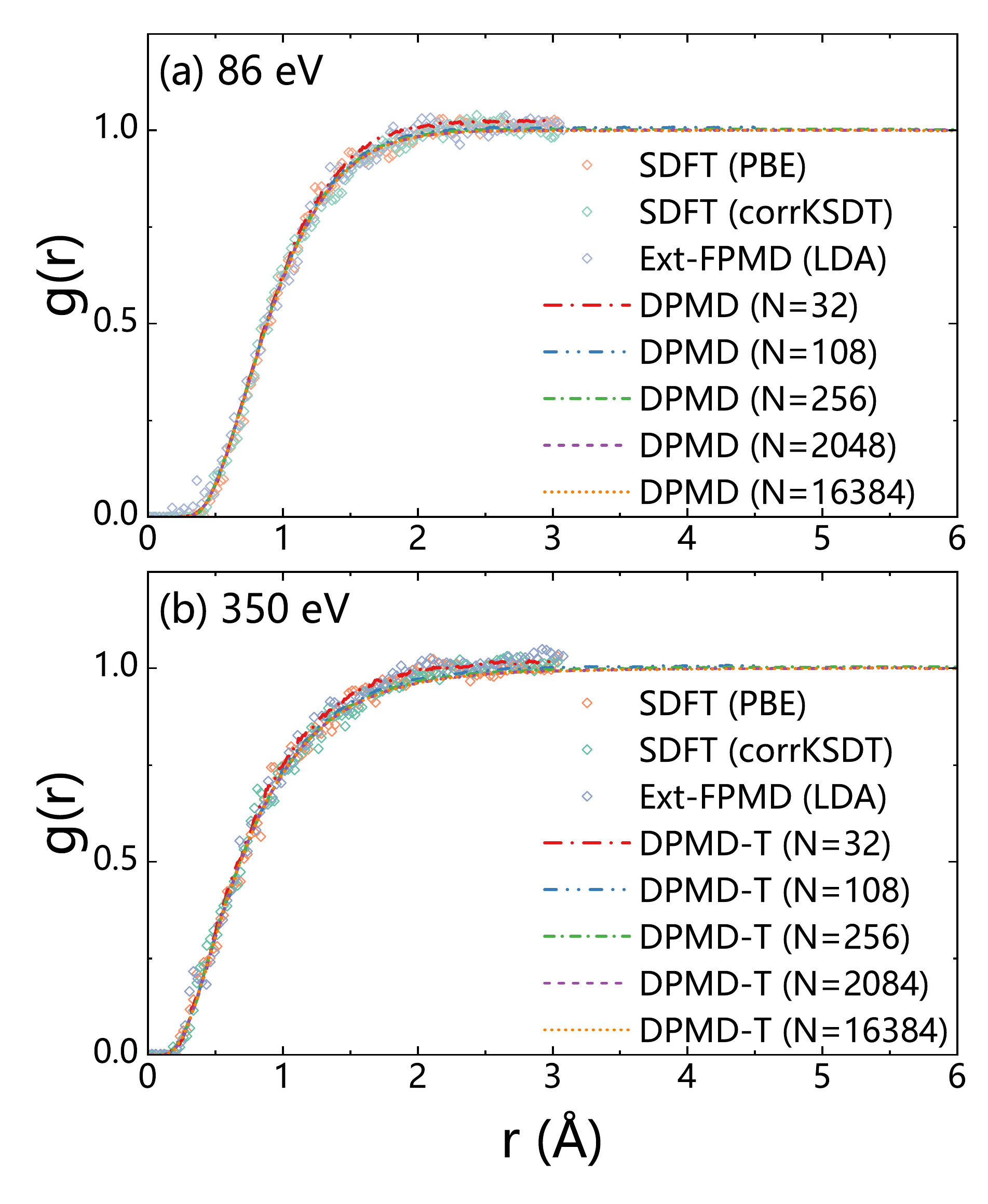}
\end{center}
\caption{
\MCS{
 (Color online) Radial distribution functions $g(r)$ of B systems with a density of 2.46 $\mathrm{g/cm^3}$ at temperatures of (a) 86 eV and (b) 350 eV.
 The $g(r)$ obtained by Ext-FPMD with the LDA functional comes from Blanchet {\it et al.}~\cite{22CPP-Blanchet}
 The SDFT calculations are performed with the PBE~\cite{96L-PBE} and corrKSDT~\cite{18L-corrKSDT} exchange-correlation functionals.
 The number of B atoms is set to 32 in the first-principles molecular dynamics simulations.
 DPMD depicts the model trained by the traditional DP method~\cite{18CPC-deepmd}, while DPMD-T means the model trained by the TDDP method~\cite{20PP-Yuzhi}. $N$ is the number of B atoms in a cell, which ranges from 32 to 16384 in the deep-potential-based simulations.}
}
\label{fig: rdf-SDFT}
\end{figure}

\MCS{
 Previous works have employed the traditional KSDFT coupling with BOMD to study WDM at relatively low temperatures. The examples include the shock Hugoniot curves~\cite{21PRE-Militzer}, the radial distribution function~\cite{12L-Driver, 20JPCM-Qianrui}, the ion-ion static structure factor~\cite{20JPCM-Qianrui}, and the ion-ion dynamic structure factor~\cite{20JPCM-Qianrui, 21PRR-DaiJiayu}, etc. However, most of these calculations are limited due to the high computational costs of dealing with electrons at extremely high temperatures. 
 In order to substantially accelerate the calculations, we choose the DPMD method to perform BOMD calculations for warm sense B systems, and the training data are obtained from efficient SDFT calculations for warm dense B at temperatures of 86 and 350 eV. Note that we use the traditional DP method~\cite{18CPC-deepmd} to train the data at a temperature of 86 eV. However, we utilize the TDDP method~\cite{20PP-Yuzhi} and include the electron temperature as an input parameter of the neural network to train the high-temperature data (350 eV), which shows a better performance than the traditional DP method at such a high temperature.
  }

\MCS{
 In detail, we perform SDFT-based BOMD simulations for a 32-atom B system with a density of 2.46 $\mathrm{g/cm^3}$.
 At the temperature of 86 eV, we adopt 
 a $\Gamma\ k$-point mesh with the number of Kohn-Sham orbitals being $N_{\mathrm{KS}}=128$ and the number of stochastic orbitals being $N_{\mathrm{sto}}=128$.
 For a higher temperature of 350 eV, a $2\times2\times2$\ shifted $k$-point mesh is used with the number of stochastic orbitals being $N_{\mathrm{sto}}=256$.
 We note that convergence with respect to the plane-wave energy cutoff and $k$-point mesh is examined to ensure the computational error of the total energy is within 0.01\%.
 The BOMD simulations are performed in the NVT ensemble with the ion temperature controlled by the velocity-rescaling thermostat.
 The electrons and ions in the system are set to the same temperature. The time step is chosen according to $\Delta{t} \sim 1 /\left(\rho^{1 / 3} T_{e}^{1 / 2}\right)$, where $T_e$ is the temperature of electrons and $\rho$ is the density. As a result, the time step is chosen to be 0.035 and 0.007 fs for simulations at 86 and 350 eV, respectively.
 In each BOMD trajectory, 4000 MD steps are performed. We then collect the atomic positions, the total energies $E$, the atomic forces $\mathbf{F}_i$ of each atom $i$, as well as the virial tensors $\Xi$ as the training data to generate DP models for B.
 \CT{Although stochastic DFT exhibits favorable scalability with increasing atoms~\cite{13L-Baer}, previous research~\cite{19PRM-Zhang, 20NC-Zeng, 20PP-Yuzhi, 23PRM-Chen} suggests that having 32 Boron atoms in the training dataset is enough to generate an accurate deep potential. The use of 32 Boron atoms to generate the training set is a choice that balances efficiency and accuracy.}
}

 \MCS{
 In DPMD simulations, we adopt the NVT ensemble with the Nosé-Hoover thermostat~\cite{84JCP-Nose, 85PRA-Hoover}. We use the LAMMPS package~\cite{95JCP-LAMMPS}. The number of B atoms ranges from 32 to 16384.
 A time step of 0.07 fs is set for the system at 86 eV and 0.01 fs for the system at 350 eV. We perform 400000 steps of DPMD simulations to yield the radial distribution functions, the static structure factors, and the dynamic structure factors. Furthermore, one million MD steps are performed to compute the shear viscosity.
 }

\MCS{
 After the BOMD trajectories are generated, the radial distribution function can be evaluated according to
 \begin{equation}
	g(r)=\frac{V}{4\pi r^2N(N-1)}\bigg\langle\sum_{i=1}^N
	\sum_{j=1,j\neq i}^{N}\delta(r-|\mathbf{r}_i-\mathbf{r}_j|)\bigg\rangle,
 \end{equation}
 where $V$ is the cell volume, $N$ is the number of atoms, $\mathbf{r}_i$ and $\mathbf{r}_j$ are atomic coordinates of atoms $i$ and $j$, and $\langle\cdots\rangle$ means the the ensemble average.
 We plot $g(r)$ of warm dense B with a density of 2.46 $\mathrm{g/cm^3}$ at 86 and 350 eV in Fig.~\ref{fig: rdf-SDFT}. We use Ext-FPMD~\cite{22CPP-Blanchet}, SDFT with two different XC functionals (PBE and corrKSDT), and DPMD methods to perform BOMD simulations. 
 We have the following findings.
 First, we find that the SDFT results are in excellent agreement with those obtained from Ext-FPMD. Second, there are no substantial differences between the PBE~\cite{96L-PBE} and the finite-temperature XC functional corrKSDT~\cite{18L-corrKSDT}, which indicates that temperature effects in the exchange-correlation functional are not significant.
 Third, as expected, the $g(r)$ is not smooth due to a limited number of MD steps (4000). However, by employing the DPMD method, we not only achieve accurate $g(r)$ that agree well with the SDFT results but also obtain a smooth $g(r)$, as a larger number of atoms (108 to 16384) and a longer trajectory (400000 steps) are considered. Importantly, size effects can be largely mitigated, as evidenced by the convergence of the $g(r)$ at around 108 atoms.
 }
 
	
	
\subsection{Ion-Ion Static Structure Factors}
	
\begin{figure}[htbp]
\begin{center}
\includegraphics[width=8.5cm]{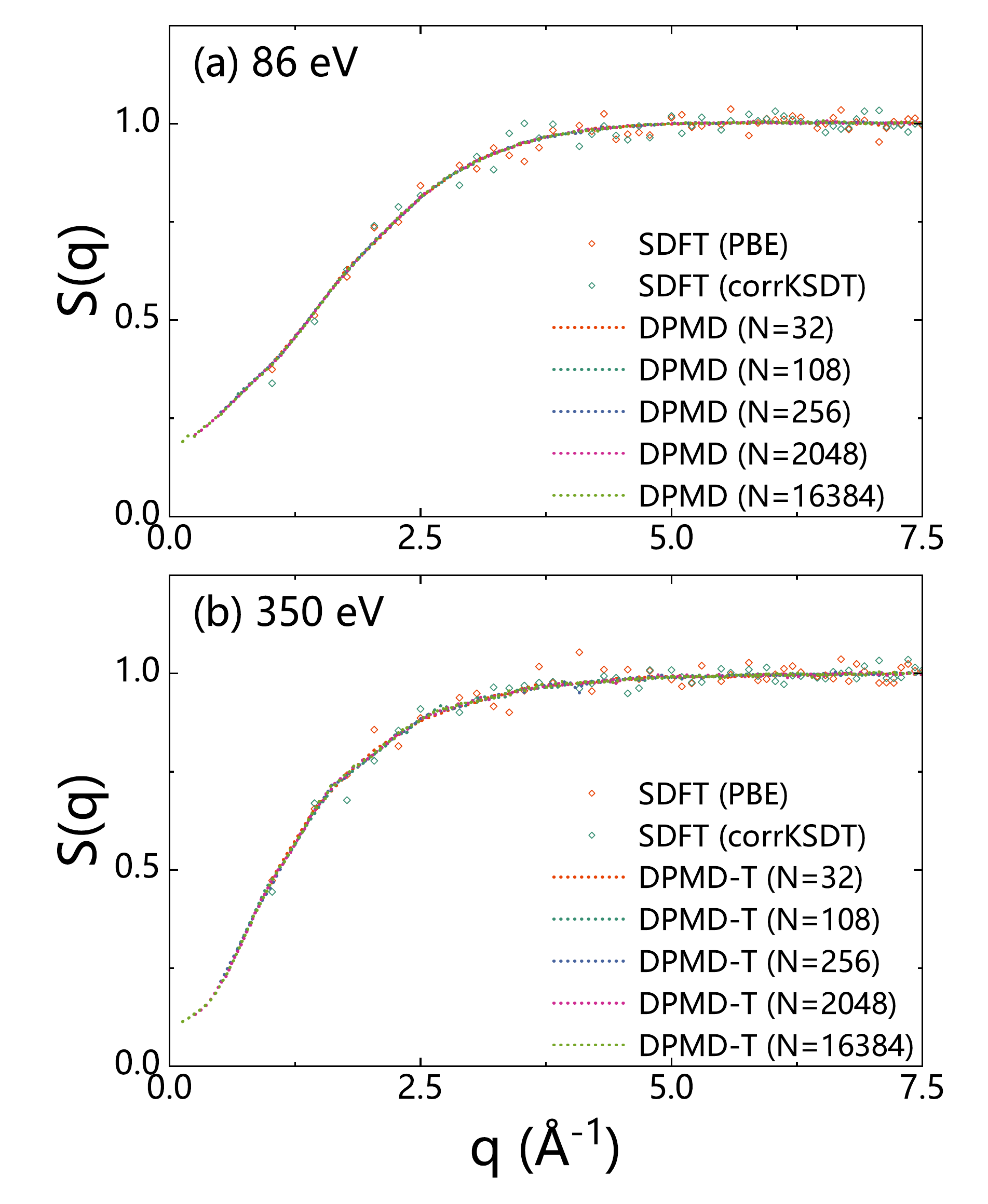}
\end{center}
\caption{
\MCS{
	(Color online) Static structure factors $S(q)$ of B with a density of 2.46 $\mathrm{g/cm^3}$ at temperatures of (a) 86 eV and (b) 350 eV.
 The SDFT calculations are performed with the PBE~\cite{96L-PBE} and corrKSDT~\cite{18L-corrKSDT} exchange-correlation functionals.
 The number of B atoms is set to 32 in the first-principles molecular dynamics simulations.
 DPMD depicts the model trained by the traditional DP method~\cite{18CPC-deepmd}, while DPMD-T indicates the model trained by the TDDP method~\cite{20PP-Yuzhi}. $N$ is the number of B atoms in a cell ranging from 32 to 16384 in the deep-potential-based simulations.}
	}
 \label{fig: ssf}
\end{figure}

\MCS{
 The ion-ion static structure factor $S(q)$ measured from diffraction experiments~\cite{71PRA-Greenfield,80B-Svensson} contains information regarding the spatial arrangement of particles in a material.
 The formula of $S(q)$ is
 \begin{equation}
 S(q)=\frac{1}{N}\left\langle\sum_{i=1}^{N} \sum_{j=1}^{N} e^{i \mathbf{q} \cdot\left(\mathbf{r}_{i}-\mathbf{r}_{j}\right)}\right\rangle,
 \end{equation}
 where $N$ is the number of atoms, while $i$, $j$ denote atoms, and $q$ is a wave vector.
 Here, we perform BOMD simulations on a 32-atom cell by SDFT with the PBE~\cite{96L-PBE} and corrKSDT~\cite{18L-corrKSDT} exchange-correlation functionals. Moreover, we employ DPMD to calculate the $S(q)$ for cells containing 32, 108, 256, 2048, and 16384 B atoms with a density of 2.46 $\mathrm{g/cm^3}$. The results for systems at 86 and 350 eV are illustrated in Figs.~\ref{fig: ssf}(a) and (b), respectively.
 It is noteworthy that the data points of $S(q)$ generated by SDFT exhibit oscillations due to the limited number of MD steps (4000). However, the DPMD simulations offer more converged results as they allow for a larger cell size with considerably more atoms and a substantially longer trajectory in BOMD simulations.
 Furthermore, with the use of larger cells in DPMD, we can obtain reasonable low-$q$ information of $S(q)$, which signifies the long-ranged order of systems.
 }
	
\subsection{Ion-Ion Dynamic Structure Factors}
	
\begin{figure}[htbp]
\begin{center}
\includegraphics[width=8.5cm]{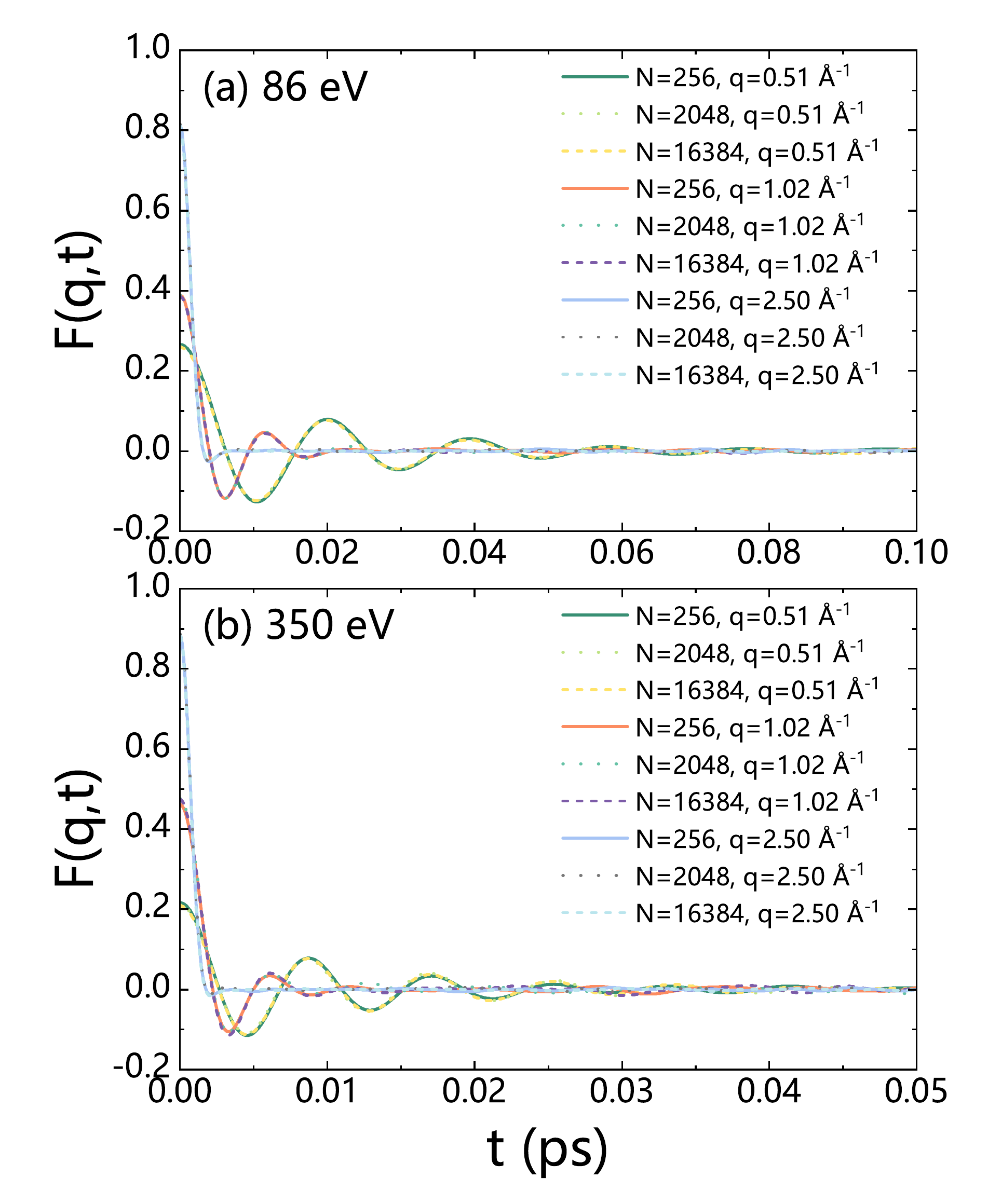}
\end{center}
\caption{
\MCS{ (Color online) Intermediate scattering functions $F(q, t)$ of B with a density of 2.46 $\mathrm{g/cm^3}$ as calculated from DPMD trajectories.
 Three system sizes (256, 2048, and 16384 atoms) are adopted in DPMD simulations at two temperatures of (a) 86 eV and (b) 350 eV. Three wave vectors are chosen, i.e., $q$=0.51, 1.02, and 2.50 \AA$^{-1}$.
 }
}
\label{fig: isf}
\end{figure}
	
\begin{figure}[htbp]
\begin{center}
\includegraphics[width=8.5cm]{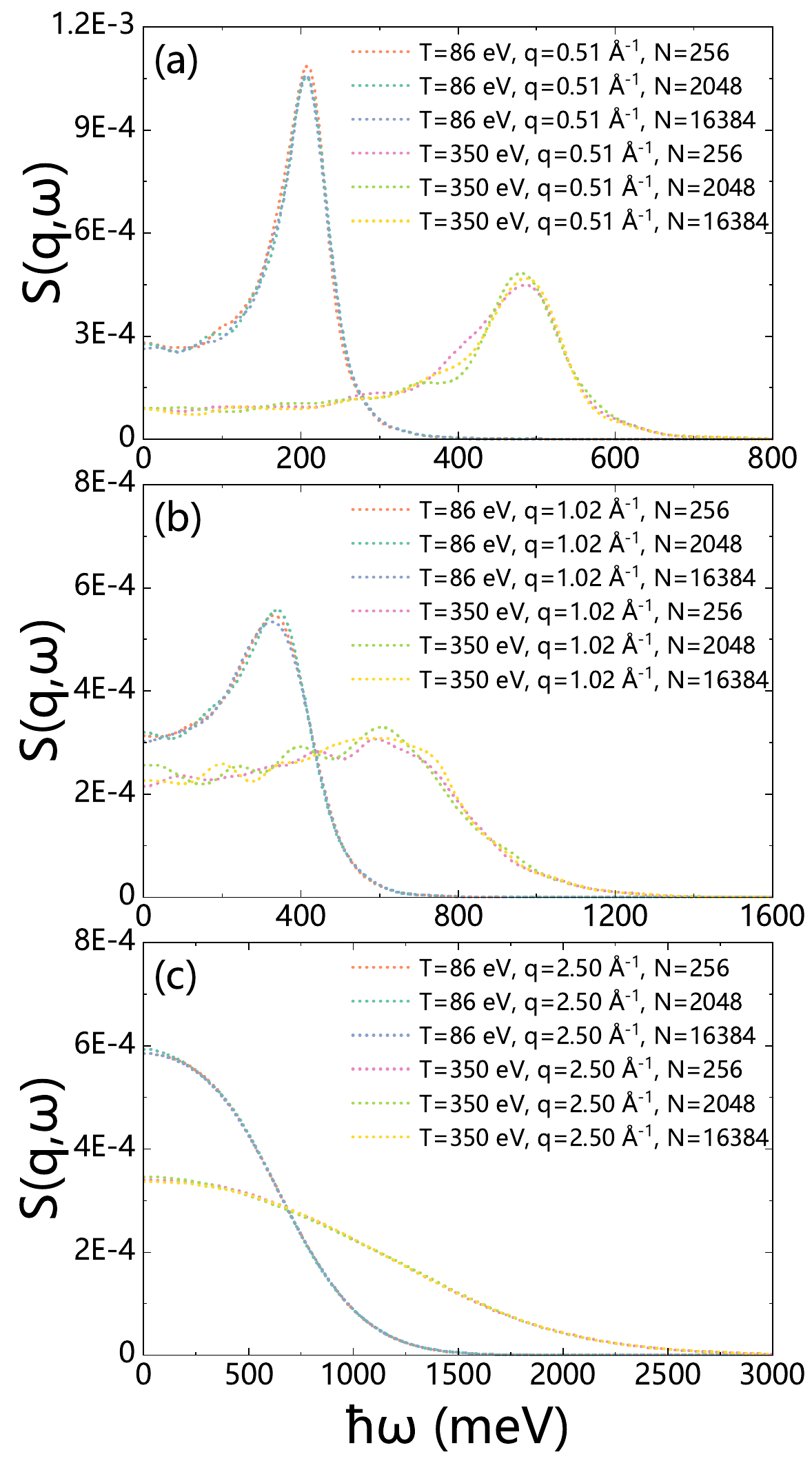}
\end{center}
	\caption{
 \MCS{
	(Color online) Ion-ion dynamic structure factors $S(q, \omega)$ of B with a density of 2.46 $\mathrm{g/cm^3}$ as computed from DPMD trajectories. Three system sizes (256, 2048, and 16384 atoms) are adopted. The wave vectors are chosen to be (a) $q$=0.51 \AA$^{-1}$, (b) 1.02 \AA$^{-1}$, and (c) 2.50 \AA$^{-1}$.}
	}
\label{fig: dsf}
\end{figure}

\MCS{
 The collective dynamics of ionic density fluctuations can be characterized by the ion-ion dynamic structure factor $S(q, \omega)$, which is experimentally measurable~\cite{19PP-Kahlert} and plays a crucial role in investigating ion dynamics, including collective modes~\cite{10PNAS-Giordano}, dissipation processes~\cite{17NC-Mabey}, and others.
 In practice, $S(q, \omega)$ can be computed from the intermediate scattering function $F(q, t)$ via Fourier transform with the formula being
 \begin{equation}
	S(q, \omega)=\frac{1}{2 \pi} \int_{-\infty}^{+\infty} F(q, t) e^{i \omega t} \mathrm{~d} t.
 \end{equation}
 Here $F(q, t)$ takes the form of
 \begin{equation}
	F(q, t)=\frac{1}{N}\langle\rho(\mathbf{q}, t) \rho(-\mathbf{q}, 0)\rangle,
 \end{equation}
 where $N$ is the number of ions, $\rho(\mathbf{q}, t)$ is defined as
 \begin{equation}
 \rho(\mathbf{q}, t)=\sum_{i=1}^{N} e^{i \mathbf{q} \cdot \mathbf{r}_{i}(t)},
 \end{equation}
 where $\mathbf{r}_{i}(t)$ is the atomic coordinates for atom $i$ at time $t$.
 }

 \MCS{
 Figs.~\ref{fig: isf}(a) and (b) illustrate the intermediate scattering function $F(q, t)$ of warm dense B at 86 and 350 eV, respectively. Three wave vectors are chosen ($q$=0.51, 1.02, and 2.50) while three sizes of cells are tested (256, 2048, and 16384 atoms). We find the 256-atom cell is large enough to converge $F(q, t)$ for both temperatures, which is beyond the capabilities of SDFT-based BOMD simulations.
 }

 Next, we obtain the ion-ion dynamics structure factors $S(q, \omega)$ of warm dense B by performing the Fourier transform of $F(q, t)$. The results associated with wave vectors $q$ being 0.51, 1.02, and 2.50 \AA$^{-1}$ are shown in Figs.~\ref{fig: dsf}(a), (b), and (c), respectively. In each figure, two temperatures (86 and 350 eV) and three system sizes (256, 2048, and 16384 atoms) are adopted.
 For the wave vector, $q$=0.51 \AA$^{-1}$, Fig.~\ref{fig: dsf}(a) shows the well-pronounced ion-acoustic modes located at $\omega$ = 206.78 meV for 86 eV and 486.80 meV for 350 eV.
 When $q$ increases to 1.02 \AA$^{-1}$ in Fig.~\ref{fig: dsf}(b), the peak of $S(q, \omega)$ becomes lower, and its location turns to 324.95 meV for 86 eV and 616.53 meV for 350 eV.
 Notably, the ion-acoustic mode $S(q, \omega)$ disappears when $q$ = 2.50 \AA$^{-1}$, because the non-collective mode dominates at large $q$.
 The above results of $S(q, \omega)$ demonstrate that the DP method can predict the long-ranged structural and time correlation of WDM.
 \CTS{For high temperatures up to hundreds of eV, there are experimental measurements of the ion-ion static structure factors and dynamic structure factors via X-ray Thomson scattering for materials, such as CH~\cite{16PRE-Kraus} and Be~\cite{23N-Doppner}. To the best of our knowledge, no experimental data are available for B at the temperatures of 86 and 350 eV.}
	
 \subsection{Shear Viscosities}
	
 \begin{figure}[htbp]
 \begin{center}
 \includegraphics[width=8.5cm]{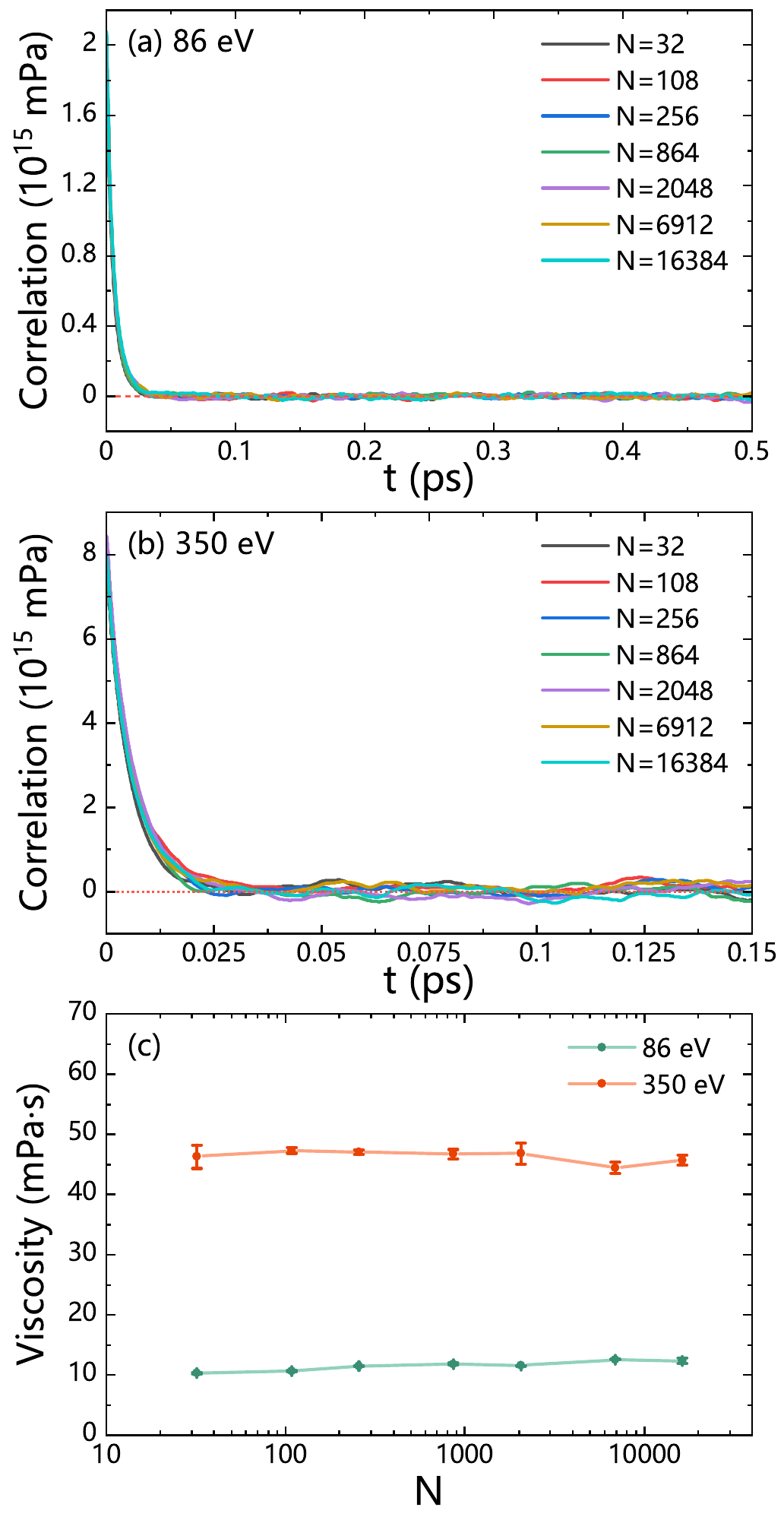}
 \end{center}
 \caption{
 \MCS{
 (Color online) Stress auto-correlation functions (Eq.~\ref{eq:autocorre_stress}) of warm dense B with a density of 2.46 $\mathrm{g/cm^3}$ at temperatures of (a) 86 eV and (b) 350 eV. (c) Shear viscosity of B. We use DPMD simulations with the cells containing 32, 108, 256, 864, 2048, 6912, and 16384 atoms. Error bars represent the standard deviations.}
 }
 \label{fig: vis}
 \end{figure}

\MCS{The shear viscosity is a crucial parameter in WDM studies, but obtaining a converged viscosity using traditional first-principles molecular dynamics is computationally expensive. However, this challenge can be significantly mitigated by employing the DPMD method with the training data from the SDFT method. One way to compute the shear viscosity $\eta$ of WDM is using the Green-Kubo relations~\cite{54JCP-Green, 57JPSJ-Kubo}, which links the shear viscosity to the integral of the stress auto-correlation function with the form of
 \begin{equation}\label{eq:autocorre_stress}
 \eta=\frac{V}{3 k_{B} T} \int_{0}^{+\infty}\left(\sum_{\alpha \beta}\left\langle P_{\alpha \beta}(0) P_{\alpha \beta}(t)\right\rangle\right) \mathrm{~d} t,
 \end{equation}
 where $V$ is the volume of the system, $T$ is the temperature, $k_B$ is the Boltzmann constant, and $P_{\alpha\beta}(t)$ ($\alpha\beta\in \{xy, xz, yz\}$) is any of the three independent off-diagonal elements of the stress tensor at time $t$. The above formula can be used when DPMD trajectories are generated with the stress tensors.}

 \MCS{
 The calculated stress auto-correlation functions of B at a density of 2.46 $\mathrm{g/cm^3}$ and temperatures of 86 and 350 eV are displayed in Figs.~\ref{fig: vis}(a) and (b), respectively. In practice, the computed shear viscosity may be affected by the system size and the trajectory length of molecular dynamics simulations.
 Therefore, we choose seven different system sizes with the number of atoms per cell ranging from 32 to 16384.
 During DPMD simulations, each system is first relaxed for 50000 steps. Next, 1 million steps of MD simulations are performed to calculate the stress auto-correlation function.
 In detail, the trajectory length is 70 ps for 86 eV and 10 ps for 350 eV.
 We take values from 0.105 to 0.305 ps (0.05 to 0.1 ps) to compute the averaged shear viscosity for the system at 86 eV (350 eV), and the predicted values are shown in Fig.~\ref{fig: vis}(c) with small error bars. As a result, the obtained shear viscosity of B varies from 10.3 to 12.3 $\mathrm{m Pa \cdot s}$ at 86 eV and from 44.4 to 47.3 $\mathrm{m Pa \cdot s}$ at 350 eV. The above results show no substantial finite-size effects for the shear viscosity of B, which is consistent with previous conclusions~\cite{15AJ-ChenMohan, 17B-ChenMohan, 22npjCM-vis}.
 }
 %

\MCS{
 There are other formulas that can predict the shear viscosity of materials.
  For example, we notice that a recent work proposes an extended random-walk shielding-potential viscosity model (ext-RWSP-VM)~\cite{22PRE-ChengYuqing, 23arX-ChengYuqing} to elevate the shear viscosity of materials in WDM and HDP states. The viscosity is evaluated by the formula of
 \begin{equation}
     \eta=\frac{\sqrt{3 m k_{B} T}}{\pi d^{4}} I,
 \end{equation}
 where $d$ is the collision diameter introduced by hard-sphere concept, and $I$ is a quantity that is relevant to $T$.
 According to the ext-RWSP-VM method, we obtain the viscosities of B to be 12.8 and 47.8 $\mathrm{m Pa \cdot s}$ for temperatures of 86 and 350 eV, respectively. More details can be found in SI. Interestingly, the data are close to our first-principles results. 
 In addition, we find the shear viscosity of plasma can also be described by the approximate formula~\cite{23-Handbook} of 
 \begin{equation}
 \label{eq: vis}
	\eta=\frac{2}{3 \sqrt{\pi}} \frac{\sqrt{m k_B T}}{\sigma_o},
 \end{equation}
 where $m$ is the atomic mass and $\sigma_o$ is the total collision cross section ($\sim 10^{-20} \mathrm{m^2}$).
 Thus, the estimated shear viscosities of B are around $\sim$ 18.7 and 37.8 $\mathrm{m Pa \cdot s}$ for temperatures of 86 and 350 eV, respectively.
 It should be noted that $\sigma_o$ in the approximate formula is assumed to be a constant; however, it is related to the relative velocity between atoms~\cite{23-Handbook}.
 As the relative velocity increases, the interaction time decreases, leading to a reduced probability of collisions occurring.
 In other words, $\sigma_o$ decreases with increasing temperature.
 We find that the calculated shear viscosities from DPMD are also consistent with the approximated values obtained using Eq.~\ref{eq: vis}.
 }

 \section{Conclusions}
 \label{Conclusions}
 
 \MCC{
 Simulating WDM with first-principles accuracy had long been challenging due to the existence of \CT{the partial occupation of a large number of high-energy KS eigenstates} and the resulting limitations in the time and space scales. Our work suggested that the advent of the SDFT method and machine-learning-based molecular dynamics can be of great help in overcoming the difficulties. The SDFT method described in this work had been implemented with the plane-wave basis set and the $k$-point sampling method, which was enabled in the ABACUS package (https://github.com/deepmodeling/abacus-develop). 
 In this work, we validated the SDFT-based BOMD method by performing a series of tests for warm dense B and C.}
   
 	
 By combining SDFT with the DP method, we substantially extended the time and space scales of simulating warm dense B and reduced the finite size effect.
 Besides, we studied the structural properties, dynamic properties, and transport coefficients, such as radial distribution functions, static structure factors, ion-ion dynamic structure factors, and shear viscosities.
 This work validated combining stochastic density functional theory with machine learning techniques to study high-temperature systems. We also offered new insights into the properties of warm dense matter. 
 \CT{In future work, we intend to explore the generation of training data with a larger number of atoms.}
 Future research may further refine these methods and expand their applicability to other materials and temperature ranges.

 \begin{acknowledgments}
 We thank Xinyu Zhang for the helpful discussions. We thank Hang Zhang for proofreading the manuscript. We thank the electronic structure team (from AI for Science Institute, Beijing) for improving the ABACUS package from various aspects. This work is supported by the National Natural Science Foundation of China under Grant Nos. 12122401 and 12074007.
 The numerical simulations were performed on the High-Performance Computing Platform of Beijing Super Cloud Computing Center and the Bohrium cloud computing platform of DP Technology Co., LTD.
\end{acknowledgments}

	\bibliography{ref}

	\end{document}


\title{Supporting Information for Combining stochastic density functional theory with deep potential molecular dynamics to study warm dense matter}

        \author{Tao Chen}
        \affiliation{HEDPS, CAPT, College of Engineering and School of Physics, Peking University, Beijing, 100871, P. R. China}

        \author{Qianrui Liu}
        \affiliation{HEDPS, CAPT, College of Engineering and School of Physics, Peking University, Beijing, 100871, P. R. China}

        \author{Yu Liu}
        \affiliation{HEDPS, CAPT, College of Engineering and School of Physics, Peking University, Beijing, 100871, P. R. China}

        \author{Liang Sun}
        \affiliation{HEDPS, CAPT, College of Engineering and School of Physics, Peking University, Beijing, 100871, P. R. China}

        \author{Mohan Chen}
        \thanks{Corresponding author. Email: mohanchen@pku.edu.cn}
        \affiliation{HEDPS, CAPT, College of Engineering and School of Physics, Peking University, Beijing, 100871, P. R. China}
        
	\date{\today}
	\maketitle

\section{Stochastic Density Functional Theory}
Figs.~\ref{fig: 86eV} and~\ref{fig: 350eV} illustrate the forces acting on each atom of B at temperatures of 86 and 350 eV, respectively. In the tests, different numbers of Kohn-Sham (KS) and stochastic orbitals, as well as different sizes of $k$ points are adopted. The density of B systems is 2.46 $\mathrm{g/cm^3}$.
Fig.~\ref{fig: force} shares the same data with Table II of the manuscript and shows additional three force components of each
atom in the 32-atom B cell from 9 independent runs of stochastic density functional theory (SDFT) with different stochastic orbitals.
\CT{Fig.~\ref{fig: md} illustrates the temperature, total energy, and pressure of molecular dynamics (MD) simulations for the temperature of 86 and 350 eV with Perdew-Burke-Ernzerhof (PBE)~\cite{96L-PBE} and corrKSDT~\cite{18L-corrKSDT}. The results indicate that the temperature, total energy, and pressure are under control when stochastic forces are used to drive a Born-Oppenheimer molecular dynamics (BOMD) trajectory. 
}

        %

\section{Viscosity}
        The viscosity in the ext-RWSP-VM~\cite{22PRE-ChengYuqing, 23arX-ChengYuqing} model is evaluated by the formula of
        \begin{equation}
        \eta=\frac{\sqrt{3 m k_{B} T}}{\pi d^{4}} I,
        \end{equation}
        where $d$ is the collision diameter introduced by hard-sphere concept, and $I$ is a quantity that is relevant to $T$.
        $I$ can be expressed as
        \begin{equation}
        I=2 r_{0}^{2} \frac{K[2(1-K)+(1+K) \ln K]}{(1+\sqrt{K})^{2}(1-K)^{2}}.
        \end{equation}
        Here
        \begin{equation}
        K=\left(\frac{r_{0}-a}{a}\right)^{2},
        \end{equation}
        where
        \begin{equation}
        a=\frac{q^{2}}{3 k_{B} T+2 q^{2} / r_{0}}
        \end{equation}
        and
        \begin{equation}
        q^{2}=\frac{(\bar{Z} e)^{2}}{4 \pi \varepsilon_{0}}.
        \end{equation}
        In the above formula, $\bar{Z}$ is the average ionization, and $r_0$ is the cutoff distance which is introduced by the Debye shielding effect.
        The ext-RWSP-VM model assumes that the cutoff distance equals the Debye length
        \begin{equation}
        r_0=\lambda_{D} = \sqrt{\frac{\varepsilon_{0} k_{B} T}{n_{e} e^{2}\left(z^*+1\right)}},
        \end{equation}
        where $\varepsilon_{0}$ is vacuum permittivity, $n_e$ is the number density of the electrons, and 
        \begin{equation}
        z^{*} = \overline{Z^{2}} / \bar{Z}.
        \end{equation}
        Here, we assume that the ionization is uniform, so $z^{*}=\bar{Z}=4.30788$ for the temperature of 86 eV and $z^{*}=4.93598$ for 350 eV. 
        The average ionization $\bar{Z}=Z\times\alpha$, where $\alpha$ is the degree of ionization calculated by Eq. (14) in the main article, and $Z$ is the nuclear charge.
        The collision diameter takes the formula of
        \begin{equation}
        d=a\left[1+\frac{R}{b_{m}} \ln \left(\frac{R+b_{m}}{a}\right)\right],
        \end{equation}
        where $R=r_0-a$, and $b_m=\sqrt{r_0^2-2r_0 a}$.
        
	\newcounter{Sfigure}
	\setcounter{Sfigure}{1}
	\renewcommand{\thefigure}{S\arabic{Sfigure}}
	\begin{figure*}[htbp]
		\begin{center}
			\includegraphics[width=17cm]{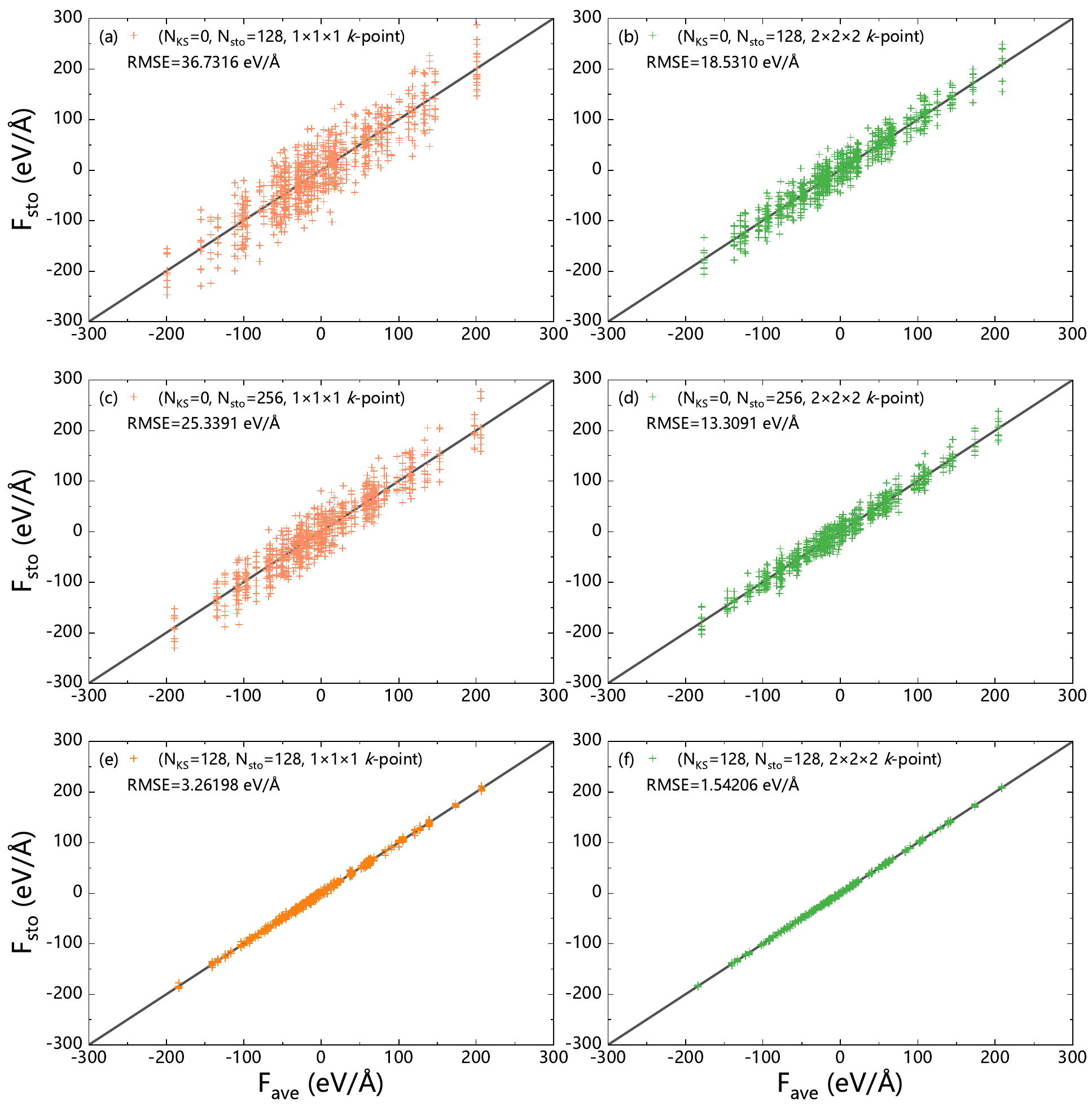}
		\end{center}
		\caption{
			(Color online) Forces on each atom of B with a density of 2.46 $\mathrm{g/cm^3}$ for 86 eV calculated by SDFT.
			$\mathrm{F_{sto}}$ is the force calculated by SDFT with nine different random seeds to generate stochastic orbitals, and their average is $\mathrm{F_{ave}}$.
			$N_{\mathrm{KS}}$ is the number of KS orbitals and $N_{\mathrm{sto}}$ is the number of stochastic orbitals.
			$1\times1\times1\ k$-point means the $\Gamma\ k$-point and $2\times2\times2\ k$-point means $2\times2\times2$ shifted $k$-point.
			RMSE is the root mean square error of forces between $\mathrm{F_{sto}}$ and $\mathrm{F_{ave}}$.
		}
		\label{fig: 86eV}
	\end{figure*}
	
	\setcounter{Sfigure}{2}
	\renewcommand{\thefigure}{S\arabic{Sfigure}}
	\begin{figure*}[htbp]
		\begin{center}
			\includegraphics[width=17cm]{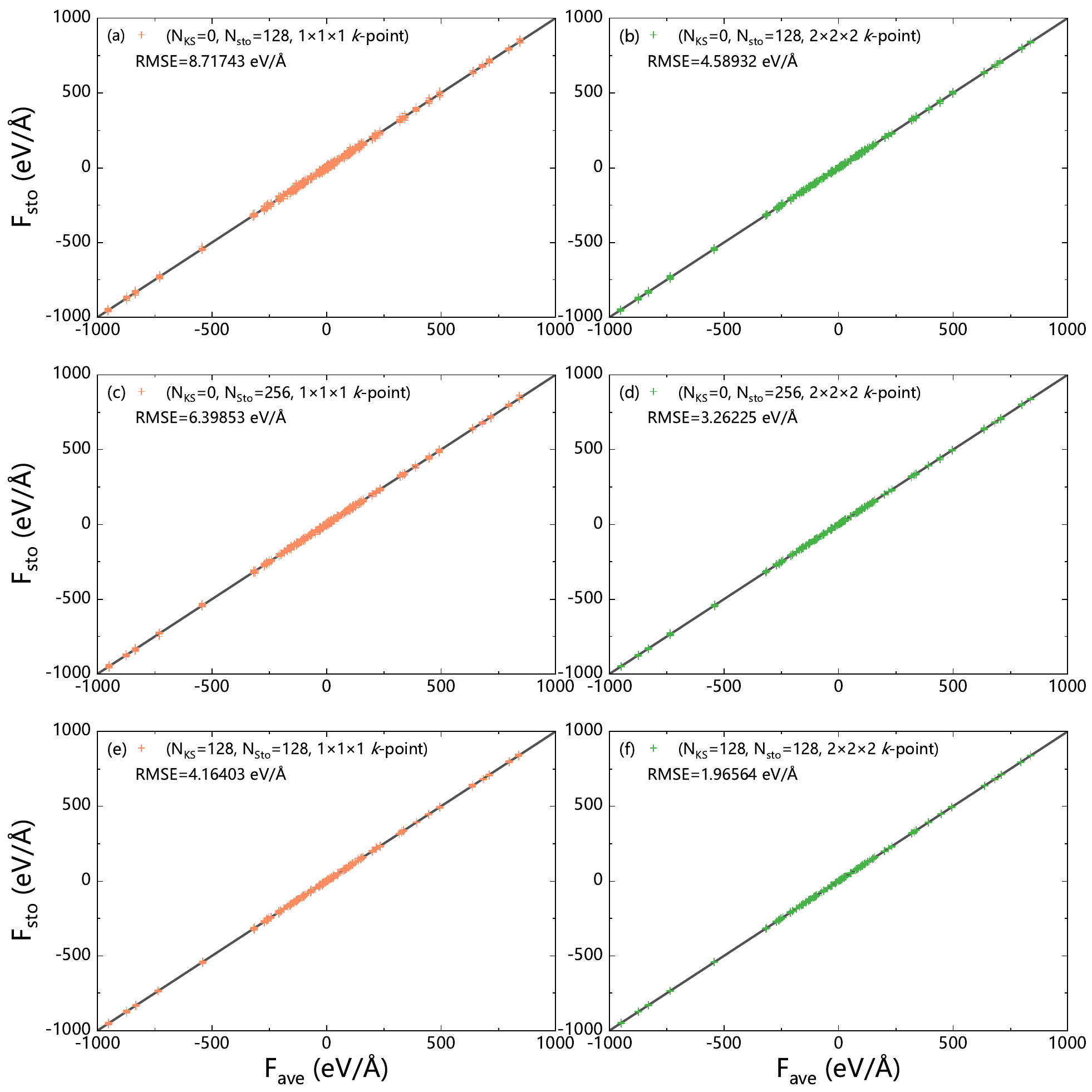}
		\end{center}
		\caption{
			(Color online) Forces on each atom of B with a density of 2.46 $\mathrm{g/cm^3}$ for 350 eV calculated by SDFT.
			$\mathrm{F_{sto}}$ is the force calculated by SDFT with nine different random seeds to generate stochastic orbitals, and their average is $\mathrm{F_{ave}}$.
			$N_{\mathrm{KS}}$ is the number of KS orbitals and $N_{\mathrm{sto}}$ is the number of stochastic orbitals.
			$1\times1\times1\ k$-point means the $\Gamma\ k$-point and $2\times2\times2\ k$-point means $2\times2\times2$ shifted $k$-point.
			RMSE is the root mean square error of forces between $\mathrm{F_{sto}}$ and $\mathrm{F_{ave}}$.
		}
		\label{fig: 350eV}
	\end{figure*}

        \setcounter{Sfigure}{3}
	\renewcommand{\thefigure}{S\arabic{Sfigure}}
	\begin{figure*}[htbp]
		\begin{center}
			\includegraphics[scale=0.85]{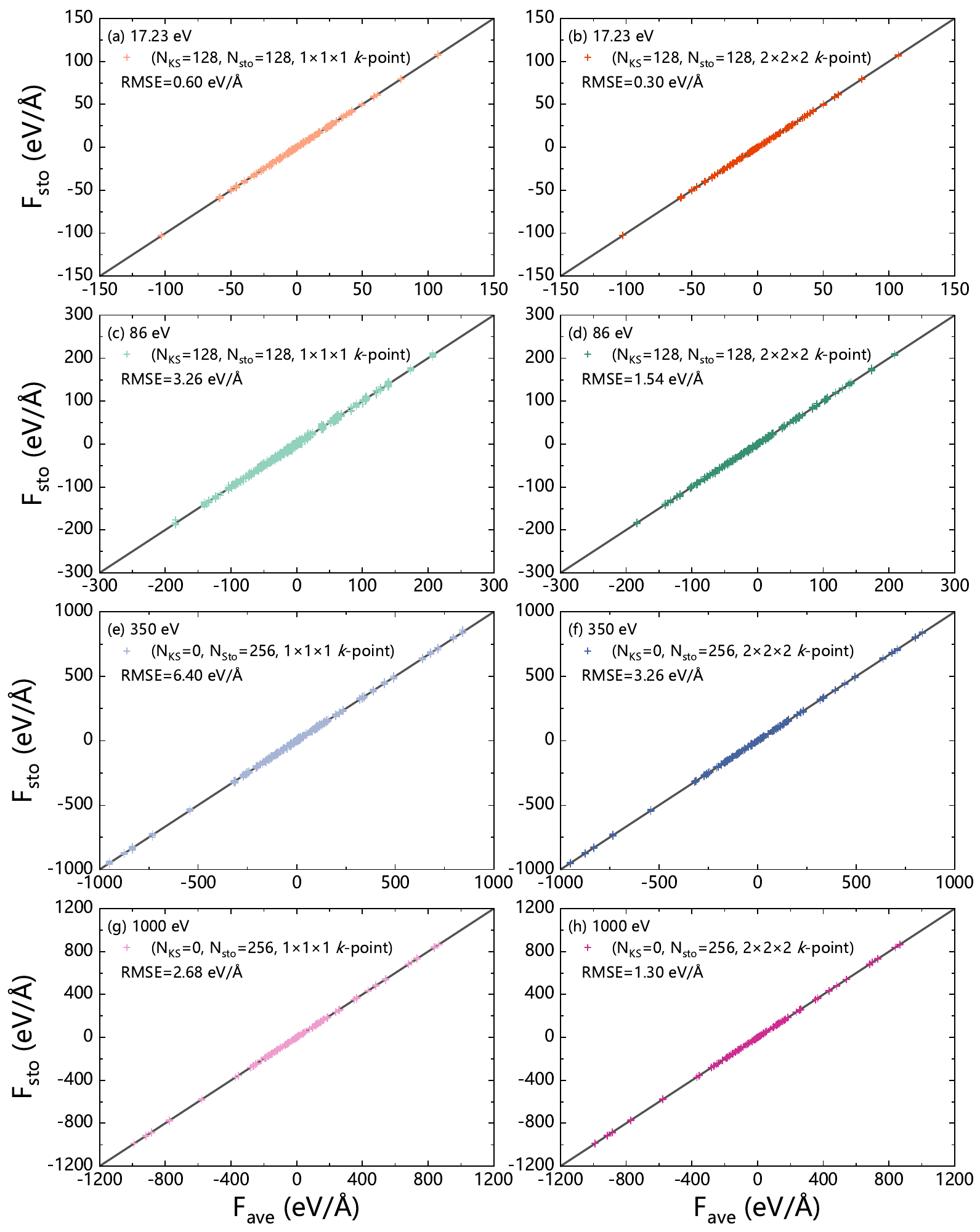}
		\end{center}
		\caption{
			(Color online) Forces acting on each B atom as obtained from 9 independent SDFT calculations with different stochastic orbitals. The B system has a density of 2.46 $\mathrm{g/cm^3}$, and the temperatures are set to 17.23 eV, 86 eV, 350 eV, and  1000 eV. For each calculation, the force acting on each atom along the $\gamma$ direction is denoted as $\mathrm{F_{sto}}$, and their average is $\mathrm{F_{ave}}$.
            $N_{\mathrm{{KS}}}$ refers to the number of KS orbitals, and $N_{\mathrm{{sto}}}$ is the number of stochastic orbitals. 
            Two sets of Monkhorst-Pack $k$-points are utilized, i.e., a $1\times1\times1\ k$-point mesh (the $\Gamma\ k$-point) and a shifted $2\times2\times2\ k$-point mesh.
			RMSE is the root mean square error of forces between $\mathrm{F_{sto}}$ and $\mathrm{F_{ave}}$.
		}
		\label{fig: force}
        \end{figure*}

        \setcounter{Sfigure}{4}
	\renewcommand{\thefigure}{S\arabic{Sfigure}}
	\begin{figure*}[htbp]
		\begin{center}
			\includegraphics[scale=0.8]{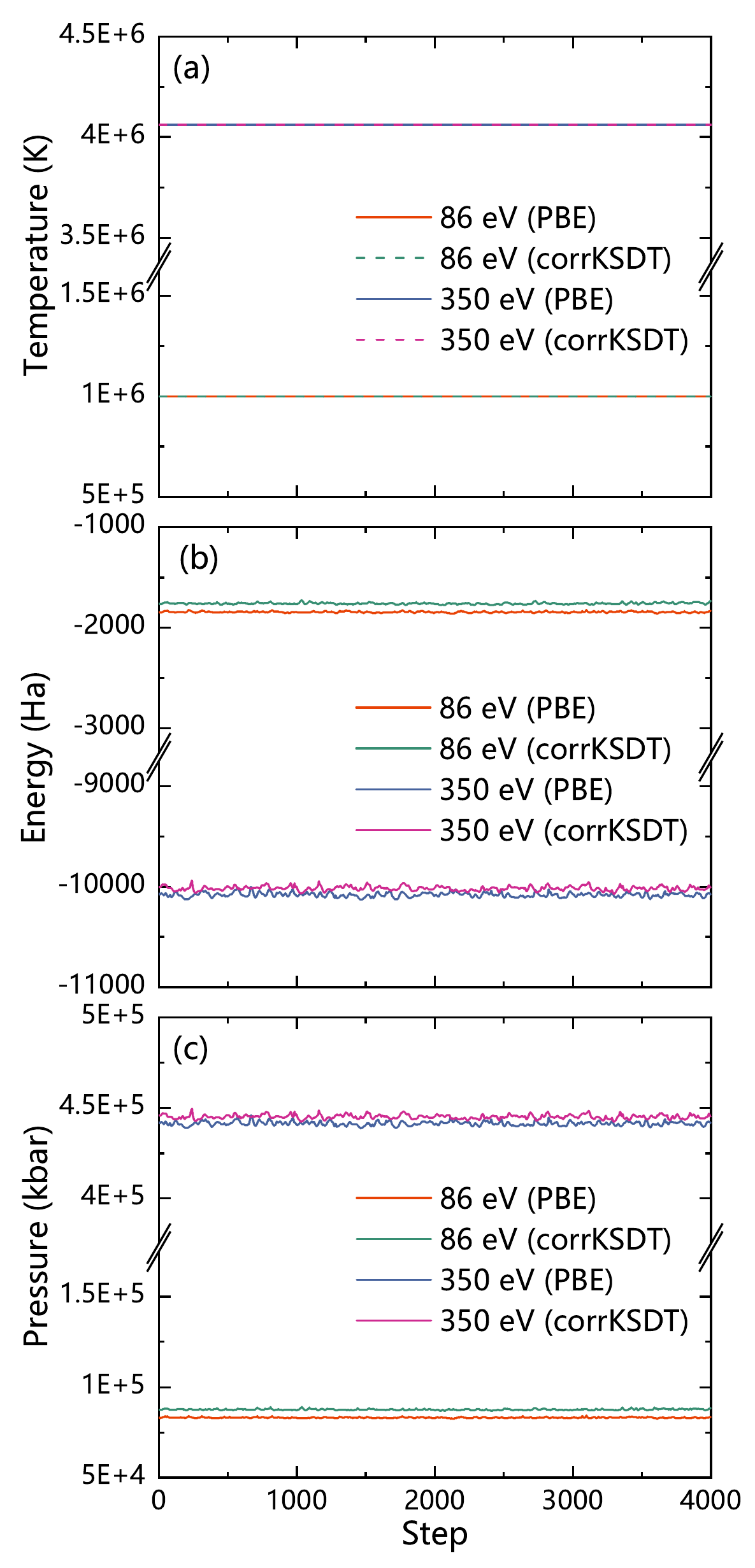}
		\end{center}
		\caption{
			\CT{(Color online) Changes of (a) temperature (in K), (b) total energy (in Ha), and (c) pressure (in kbar) with respect to MD steps when simulating B systems at temperatures of 86 and 350 eV with the PBE~\cite{96L-PBE} and corrKSDT~\cite{18L-corrKSDT} exchange-correlation functionals.
		}}
		\label{fig: md}
        \end{figure*}

        \bibliography{ref}